\documentclass[12pt]{article}
\normalbaselines \topmargin -0.25 truein \oddsidemargin 0.30
truein \evensidemargin 0.30 truein 

\begin{document}

%%%%%%%%%%%%%%%%%%%%%%%%%%%%%%%%%%%%%%%%%%%%%%%%%%%%%%%%%%%%%%
\indent \vskip 1cm \centerline{\bf MAGNETIC RELAXATION IN THE
BIANCHI-I UNIVERSE}

\vskip 3cm

%%%%%%%%%%%%%%%%%%%%%%%%%%%%%%%%%%%%%%%%%%%%%%%%%%%%%%%%%%%%%%%
\centerline{\bf Alexander Balakin {\it \footnote{e-mail:
Alexander.Balakin@ksu.ru}} } \vskip 0.3cm

\centerline{Department of General Relativity and Gravitation,}

\centerline{\it Kazan State University, Kremlevskaya street 18,
420008, Kazan, Russia}

\vskip 3cm

\date{\today}
%\maketitle

\begin{abstract}
Extended Einstein-Maxwell model and its application to the problem
of evolution of magnetized Bianchi-I Universe are considered. The
evolution of medium magnetization is governed by a relaxation type
extended constitutive equation. The series of exact solutions to
the extended master equations is obtained and discussed. The
anisotropic expansion of the Bianchi-I Universe is shown to become
non-monotonic (accelerated/decelerated) in both principal
directions (along the magnetic field and orthogonal to it). A
specific type of expansion, the so-called evolution with hidden
magnetic field, is shown to appear when the magnetization
effectively screens the magnetic field and the latter disappears
from the equations for gravitational field.
\end{abstract}

\vspace{0.8cm}

Key words: {\it anisotropic medium, polarization, magnetization, Einstein-Maxwell theory,
extended thermodynamics, extended constitutive equations}.

PACS numbers: 04.40.-b , 04.40.Nr , 04.20.Jb , 98.80.Jk

\vspace{0.5cm}

\newpage

\section{Introduction}

Observational cosmology and astrophysics indicate that many
subsystems of the Universe, which have different length-scales:
planets, stars, galaxies, clusters and superclusters of galaxies,
possess inherent magnetic field (see, e.g., the reviews
\cite{MF1,MF2,MF3} and references therein). The magnetic field
interacts with material environment and this results in the
magnetization of the latter \cite{EM}. According to basic
Einstein's ideas, the stress and energy of all three
constituents: medium, magnetic field and magnetization, - act as
sources of gravity field. It seems to be reasonable from physical
point of view, that for majority of astrophysical and
cosmological objects the contribution of a material medium, as a
source of gravity field, dominates over the contributions of
magnetic field and magnetization, nevertheless, the stress and
energy of magnetization itself can be comparable with those of
pure magnetic field. In other words, when we deal with exactly
integrable models, for which the magnetic field is considered as
a non-negligible source of gravity field in comparison with the
contribution of the material medium, one should also take into
account the magnetization as a cross-effect.

Cosmological Bianchi-I model seems to be the most convenient for
testing this idea due to its three specific features. First, this
model is non-stationary, all three constituents, medium, magnetic
field and magnetization, can evolve with different rates, thus
displaying their dynamic non-equivalence. Second, the Bianchi-I
model belongs to the class of the exactly integrable ones, thus
providing the analysis of singular behaviour of the model. Third,
the Bianchi-I model is spatially anisotropic and thus admits a
self-consistent description of the uniaxial configuration of the
dynamic system containing interacting matter, magnetic field and
magnetization. There is a number of anisotropic cosmological
models, in which magnetic field is accompanied by perfect fluid
(see, e.g., \cite{ExactSolutions,dina} and references therein),
or by non-equilibrium (viscous) cosmic fluid (see, e.g.,
\cite{Tsagas,Pradhan1,Saha}). If in addition to matter and
magnetic field the third ``player", namely the magnetization,
appears in a cosmological dynamics, one can expect that the rate
of evolution of the Universe modifies. Our expectations are
motivated by the analogy with dissipative phenomena in cosmology,
described in the framework of causal (extended) thermodynamics
\cite{Israel2} - \cite{Zimdahl96}. In this theory the extended
constitutive law for the viscous fluid contains a time parameter,
which is known to introduce a specific time scale into expansion
rate. Since the evolution of magnetization has usually a
character of relaxation, a new time parameter, say relaxation
time, $\tau$, also must appear and establish a new time scale.
The interplay between $\tau$ and expansion rate parameter(s) can
introduce a qualitatively new aspects into cosmological dynamics.

In order to formulate a self-consistent Einstein-Maxwell model,
taking into account the polarization and magnetization of a {\it
non-stationary} material medium, two key elements are necessary.
The first one is an adequate energy-momentum tensor of the
electromagnetically active medium, which forms a source term in
the right-hand-side of the Einstein equations. Here we derive
explicitly such a tensor using Lagrangian formalism for the
stationary non-conducting medium with uniaxial symmetry in case
when magneto-electric cross effects are absent. Expressed in terms
of induction tensor and Maxwell tensor, this quantity happens to
be a symmetrized Minkowski stress-energy tensor (see, e.g.,
\cite{Minkowski}-\cite{Brevik1}
%\cite{Minkowski,IsraelT,Maugin78,Ginzburg,Brevik1}
for a review, historical details and terminology). When a medium
is non-stationary, the required effective stress-energy tensor is
assumed to have the same formal structure. Nevertheless, the
induction tensor acquires now a new sense: it is a sum of the
Maxwell tensor and of a polarization-magnetization tensor, the
latter quantity being considered as a new dynamic variable. The
second key element of non-stationary model is the {\it extended
constitutive equations}, which establish relations between
polarization-magnetization of a medium and electromagnetic field
strength. We use here the simplest extended constitutive
equations of a relaxation type, which are formulated
phenomenologically, based on the well-known analogs from causal
(extended) thermodynamics \cite{Israel2} - \cite{Zimdahl96}. The
main goal of this paper is an application of the formulated
extended Einstein-Maxwell (EEM) model to the description of the
magnetization dynamics in the Universe, considering the
corresponding master equations in the context of Bianchi-I
anisotropic cosmological model.

The paper is organized as follows. In Section 2 we briefly discuss
the principal details of the EEM- model. Particularly, we
introduce the effective stress-energy tensor describing the
electromagnetic field and the polarization - magnetization in a
medium, as well as we introduce the constitutive equations of a
relaxation type. In Section 3 we adopt the EEM model for the
symmetry related to Bianchi-I cosmological model and discuss the
reduced Maxwell, Einstein and constitutive equations. In Section 4
we consider exact solutions of the dynamic equation for the
magnetization with variable relaxation time parameter, discuss
general properties of these solutions and some interesting
particular cases. In Section 5 we obtain exact solutions of the
Einstein equations for the case of variable relaxation time
parameter. We distinguish two principal submodels in this context.
The first submodel describes a paramagnetic/diamagnetic dust
(Subsection 5.1) and contains three exactly integrable particular
cases. The second submodel describes the so-called longitudinal
quasi-vacuum (Subsection 5.2) and also contains three exactly
integrable particular cases. In Section 6 we obtain exact
solutions of the EEM- model for the case of constant relaxation
parameter. In Subsection 6.1 we establish a law of the
magnetization evolution. In Subsections 6.2 we consider the model
with hidden induction. In Subsection 6.3 three particular cases of
exact solutions of the EEM model, describing the submodel with
vanishing total longitudinal pressure, are discussed. In
Subsection 6.4 the example of cosmological dynamics with
non-homogeneous  non-linear equation of state of the magnetized
matter is studied. Discussions form Section 7. Appendix contains
the description of the procedure of variation of the tetrad
four-vectors, which is used in Section 2.

\section{Extended Einstein-Maxwell model}

\subsection{Stationary media with uniaxial spatial symmetry}

Let us consider a preliminary model with the following action
functional
\begin{equation}
S = \int d^4 x \sqrt{-g} \left\{ \frac{R+ 2 \Lambda}{2\kappa} +
L_{({\rm matter})} + \frac{1}{4} C^{ikmn} F_{ik} F_{mn} \right\}
\,, \label{lagran}
\end{equation}
as a {\it hint} for construction of the extended Einstein-Maxwell
model. Here $R$ is the Ricci tensor, $g$ is the determinant of
$g_{ik}$, $\Lambda$ is cosmological constant, $\kappa =8\pi G$,
$G$ is the gravitational constant. The quantity $F_{mn}$ is the
Maxwell tensor, $F_{mn}=\partial_m A_n -\partial_n A_m$, $A_m$ is
a potential four-vector of the electromagnetic field. $L_{({\rm
matter})}$ is the pure medium contribution to the Lagrangian, we
assume that this scalar does not depend on $F_{mn}$. The quantity
$C^{ikmn}$ is the linear response tensor, which describes the
influence of matter on the electromagnetic field. This tensor is
assumed to possess the following symmetries
\begin{equation}
 C^{ikmn} = - C^{kimn} = - C^{iknm} = C^{mnik}
\,, \label{eldacoeff}
\end{equation}
and is a function of the metric $g_{mn}$, time-like velocity
four-vector of the medium as a whole, $U^k$, and some space-like
vector $X^k$, pointing the privilege direction in the medium.
These four-vectors are orthogonal and normalized by unity, i.e.,
\begin{equation}
g_{pq}U^pX^q=0 \,, \quad g_{pq}U^pU^q=1 \,, \quad g_{pq}X^pX^q= -
1 \,. \label{norma}
\end{equation}
When $C^{ikmn}$ contains the Riemann tensor, the Ricci tensor and
the Ricci scalar, we deal with {\it non-minimal} Einstein -
Maxwell theory \cite{BL05,BZ05}. When $C^{ikmn}$ includes the
covariant derivative of the velocity four-vector $\nabla_i U_k$,
the corresponding model describes dynamo-optical effects
\cite{AlBa}. It is worth stressing that here we restrict ourselves
by the case when $C^{ikmn} = C^{ikmn}[g_{pq},U^k,X^l]$. The
procedure of phenomenological reconstruction of the material
tensor $C^{ikmn}$ is well-known \cite{EM,Tsypkin}. In order to
obtain $C^{ikmn}$ as a function of $g_{mn}$, $U^k$ and $X^k$ for
the medium with uniaxial symmetry one uses, first, the standard
decomposition
\begin{equation}
C^{ikmn} {=} \frac12 U^qU^s \left\{ \delta^{ik}_{pq}
\delta^{mn}_{ls}\varepsilon^{lp} {-} \epsilon^{ik}_{\ \ pq}
\epsilon^{mn}_{\ \ \ ls} (\mu^{{-}1})^{lp} {+} \nu^{lp} \left[
\epsilon^{ik}_{\ \ ls} \delta^{mn}_{pq} {+} \epsilon^{mn}_{\ \ \
ls} \delta^{ik}_{pq} \right] \right\}\,, \label{Cdecomp}
\end{equation}
where dielectric permittivity tensor, $\varepsilon^{lp}$, magnetic
impermeability tensor,  $(\mu^{-1})^{lp}$,  and magneto-electric
tensor $\nu^{lp}$ are used, defined as
\begin{equation}
\varepsilon^{pl} = 2 C^{pkln} U_k U_n \,, \quad (\mu^{-1})^{pq} =
- \frac{1}{2} \eta^p_{\ ik} C^{ikmn} \eta_{mn}^{\ \ \ q} \,, \quad
\nu^{pl} = \eta^p_{\ ik} C^{ikln} U_n \,.
 \label{emunu}
\end{equation}
The quantity $\delta^{ik}_{pq}$ is the Kronecker tensor,
$\epsilon^{ikjs}$ is the Levi-Civita tensor
\begin{equation}
\delta^{ik}_{pq} = \delta^{i}_{p}\delta^{k}_{q} -
\delta^{i}_{q}\delta^{k}_{p} \,, \quad \eta^{ikj} \equiv
\epsilon^{ikjs}U_s \,, \quad \epsilon^{ikjs} \equiv
\frac{E^{ikjs}}{\sqrt{-g}} \,, \label{eta}
\end{equation}
$E^{ikjs}$ is the completely skew - symmetric Levi-Civita symbol
with $E^{0123}=1$. The Levi-Civita tensor provides the dualization
procedure: $F^{*}_{ik} \equiv \frac{1}{2} \epsilon_{ikmn}F^{mn}$.
The second step of reconstruction of $C^{ikmn}$ tensor for the
case of uniaxial symmetry is a phenomenological representation of
the tensors $\varepsilon^{lp}$, $(\mu^{-1})^{lp}$ and $\nu^{lp}$.
In the simplest case, when the magnetoelectric cross-terms are
absent ($\nu^{lp}=0$), such relations are
\begin{equation}
\varepsilon^{lp} = \varepsilon_{\bot} \Delta^{lp}  + \left(
\varepsilon_{\bot} - \varepsilon_{||} \right) X^l X^p  \,, \quad
(\mu^{-1})^{lp} = \frac{1}{\mu_{\bot}} \Delta^{lp} + \left(
\frac{1}{\mu_{\bot}} - \frac{1}{\mu_{||}} \right) X^l X^p \,.
\label{emunuUNIAX}
\end{equation}
Here the scalar quantities $\varepsilon_{\bot}$ and
$\varepsilon_{||}$ are transversal and longitudinal coefficients
of dielectric permittivity, respectively, scalars $\mu_{\bot}$ and
$\mu_{||}$ represent coefficients of transversal and longitudinal
magnetic permeability, respectively, $\Delta^{lp} \equiv
g^{lp}-U^lU^p$ is a projector. Finally, (\ref{Cdecomp}) can be
rewritten as
$$
C^{ikmn}= \frac{1}{2\mu_{\bot}} \left[ \left(g^{im} g^{kn} -
g^{in}g^{km}\right) + \left(\varepsilon_{\bot} \mu_{\bot}-1
\right) \delta^{ik}_{pq} \ \delta^{mn}_{ls} \ g^{pl} \ U^q \ U^s
\right] +
$$
\begin{equation}
+ \frac{1}{2} U^q U^s X^l X^p \left[ \left( \varepsilon_{\bot} -
\varepsilon_{||} \right) \delta^{ik}_{pq} \ \delta^{mn}_{ls} -
\left( \frac{1}{\mu_{\bot}} - \frac{1}{\mu_{||}} \right)
\epsilon^{ik}_{\ \ pq} \epsilon^{mn}_{\ \ \ ls}\right] \,.
\label{Cfinal}
\end{equation}
When $\varepsilon_{\bot} = \varepsilon_{||} = \varepsilon$ and
$\mu_{\bot}=\mu_{||}= \mu$, the obtained tensor of material
coefficients $C^{ikmn}$ covers the well-known isotropic one (see,
e.g., \cite{EM,Tsypkin}).

\subsubsection{Maxwell equations}

Variation of the action functional (\ref{lagran}) with respect to
the four-vector of electromagnetic potential $A_i$ gives the
Maxwell equations with vanishing current of free charges
\begin{equation}
\nabla_k \left( C^{ikmn} F_{mn} \right) = 0 \,. \label{maxwell}
\end{equation}
In this case the induction tensor, $H^{ik}$, is equal to the
expression in the parentheses.

\subsubsection{Equations for gravity field}

Variation of the action functional (\ref{lagran}) with respect to
metric $g_{ik}$ yields
\begin{equation}
R^{ik} - \frac{1}{2} g^{ik} R = \Lambda g^{ik} + \kappa
T^{ik}_{({\rm total})} \,, \quad T_{({\rm total})}^{ik} = T_{({\rm
matter})}^{ik} + T_{({\rm eff})}^{ik} \,. \label{Ein}
\end{equation}
The symmetric stress - energy tensor of the material medium
$T_{({\rm matter})}^{ik}$, defined as
\begin{equation}
T^{ik}_{({\rm matter})} \equiv -
\frac{2}{\sqrt{-g}}\frac{\delta}{\delta g_{ik}} \left( \sqrt{-g}
L_{({\rm matter})} \right) \,, \label{Teff}
\end{equation}
can be written in the standard form
\begin{equation}
T^{ik}_{({\rm matter})} = W U^i U^k + q^i U^k + q^k U^i - P
\Delta^{ik} + \Pi^{ik} \,, \label{Tmatter}
\end{equation}
where $W$ is an energy density scalar of the matter, $q^i$ is a
heat-flux four-vector, $P$ is the Pascal pressure and $\Pi^{ik}$
is an anisotropic pressure tensor. The effective stress-energy
tensor has the form
\begin{equation}
T_{({\rm eff})}^{ik} = -
\frac{1}{2\sqrt{-g}}F_{pq}F_{mn}\frac{\delta}{\delta g_{ik}}
\left( \sqrt{-g} C^{pqmn} \right) \,, \label{Tem}
\end{equation}
it can be calculated directly for the medium with uniaxial
symmetry  using the decomposition (\ref{Cfinal}) and the following
formulas for the variations $\delta U^l$ and $\delta X^s$:
\begin{equation}
\delta U^l = \frac{1}{4} \delta g_{ik} \left(U^i g^{lk} + U^k
g^{li} \right) \,, \quad \delta X^s = \frac{1}{4} \delta g_{ik}
\left(X^i g^{sk} + X^k g^{si} \right) \,.\label{deltaX}
\end{equation}
The grounds of the formulas (\ref{deltaX}) are presented in the
Appendix. The variation procedure yields
\begin{equation}
T^{ik}_{({\rm eff})} \equiv  \frac{1}{4} g^{ik} C^{pqmn} F_{pq}
F_{mn} - \frac{1}{2} (C^{impq} F^k_{ \ m} +  C^{kmpq} F^i_{ \ m}
)F_{pq} \,, \label{BPisotrop}
\end{equation}
i.e., the effective stress-energy tensor is explicitly symmetric
and traceless.

\subsubsection{Resume}

When the non-conducting medium with uni-axial symmetry is the {\it
stationary} one, the Einstein-Maxwell model consists of three
ingredients:

\noindent {\it (i) Maxwell equations}

\begin{equation}
\nabla_k H^{ik} = 0 \,, \quad \nabla_k F^{*ik} = 0 \,,
\label{maxwellR}
\end{equation}

\noindent {\it (ii) constitutive equations}

\begin{equation}
H^{ik} = C^{ikmn} F_{mn} \,, \label{CE1}
\end{equation}

\noindent {\it (iii) gravity field equations}

$$
R^{ik} - \frac{1}{2} g^{ik} R = \Lambda g^{ik} + \kappa \left[ W
U^i U^k + q^i U^k + q^k U^i - P \Delta^{ik} + \Pi^{ik} \right] +
$$
\begin{equation}
+ \kappa \left[\frac{1}{4} g^{ik} H_{mn} F^{mn} - \frac{1}{2}
(H^{im} F^k_{ \ m} +  H^{km} F^i_{ \ m} ) \right] \,. \label{Ein2}
\end{equation}
It is the exact result, which follows from the variation
procedure. The electromagnetic part of the total stress-energy
tensor in the presented form coincides with the symmetrized
Minkowski electromagnetic energy-momentum tensor (see, e.g., the
review \cite{Brevik1}). It is manifestly symmetric, traceless and
does not depend explicitly on the choice of $U^i$.

\subsubsection{Extended Einstein-Maxwell model}

When the medium is non-stationary, the Einstein-Maxwell model
should be modified accordingly, the stationary equations
(\ref{maxwellR})-(\ref{Ein2}) can be used as a hint. Our {\it
ansatz} for such extension is the following: we consider the
electrodynamic equations in the same form (\ref{maxwellR}), but
express the induction tensor in terms of
polarization-magnetization tensor $M^{ik}$; instead of stationary
constitutive equations (\ref{CE1}) we introduce extended
constitutive equation of a relaxation type for the tensor of
polarization-magnetization; the gravity field equations have the
same form (\ref{Ein2}), but $H^{ik}$ is replaced by
$F^{ik}+M^{ik}$, i.e, the basic set of master equations for the
Extended Einstein-Maxwell model is
\begin{equation}
\nabla_k \left(F^{ik}+M^{ik}\right) = 0 \,, \quad \nabla_k F^{*ik}
= 0 \,, \label{maxwellRR}
\end{equation}
\begin{equation}
\tau DM^{ik} + M^{ik} =  \chi^{ikmn} F_{mn} \,, \label{dM}
\end{equation}
$$
R^{ik} - \frac{1}{2} g^{ik} R = \Lambda g^{ik} + \kappa \left[ W
U^i U^k + q^i U^k + q^k U^i - P \Delta^{ik} + \Pi^{ik} \right] +
$$
\begin{equation}
+ \kappa \left[\frac{1}{4} g^{ik} F_{mn} F^{mn} - F^{im} F^k_{ \
m} \right] + \kappa \left[\frac{1}{4} g^{ik} M_{mn} F^{mn} -
\frac{1}{2} (M^{im} F^k_{ \ m} +  M^{km} F^i_{ \ m} ) \right] \,.
\label{Ein2R}
\end{equation}
Here $\tau$ is a relaxation time and $D \equiv U^k \nabla_k$ is a
convective derivative, $\chi^{ikmn}$ is a linear susceptibility
tensor, which can be expressed in terms of $C^{ikmn}$ as
\begin{equation}
\chi^{ikmn} \equiv C^{ikmn} - \frac{1}{2} \left( g^{im}g^{kn} -
g^{in}g^{km} \right) \,. \label{CE11}
\end{equation}
The quantity $M^{ik}$ is considered in this context as a new
variable. Such extension of the constitutive equations can be
regarded as phenomenologically motivated, when the well-known
analogs from rheology  and extended thermodynamics \cite{Israel2}
- \cite{Zimdahl96} are taken into account. Thus, the novelty of
the presented model is connected, first, with the extended
constitutive equation (\ref{dM}), second, with the modified
electromagnetic source in the gravity field equations (see the
last line in (\ref{Ein2R})).

\section{Bianchi-I Universe}

Consider the rotationally isotropic Bianchi-I cosmological model
\cite{ExactSolutions,dina} with the line element
\begin{equation}
ds^2 = dt^2 - a^2(t) \left[ (dx^1)^2 + (dx^2)^2 \right] - c^2(t) \
(dx^3)^2 \,. \label{metric}
\end{equation}
We suppose that the velocity four-vector of a matter, $U^k$, has
the form $U^k=\delta^k_t$ and the space-like vector $X^i$ is $X^i=
\delta^i_3 / c(t)$. They satisfy the relations $DU^i=0$ and
$DX^i=0$. The symmetry of the model prescribes the non-diagonal
components of the total stress-energy tensor to vanish. In this
sense to make the model self-consistent one can consider, for
instance, the case when the magnetic field four-vector and the
magnetization four-vector are parallel to the $x^3$ axis. For such
a configuration of the electromagnetic field the total
stress-energy tensor, $T^{ik}_{({\rm total})}$, has four
non-vanishing components:
$$
T^0_{0({\rm total}) } \equiv {\cal W} = W + {\cal X} \,,  \quad
T^1_{1({\rm total}) } = T^2_{2({\rm total}) } \equiv - {\cal
P}_{({\rm tr})} = - (P_{({\rm tr})} + {\cal X}) \,,
$$
\begin{equation}
T^3_{3({\rm total}) } \equiv - {\cal P}_{||} = - ( P_{||} - {\cal
X}) \,, \quad  {\cal X} \equiv \frac{1}{2} H^{12} F_{12} =
\frac{1}{2} [ F^{12} F_{12} + M^{12} F_{12} ] \,. \label{Xdef}
\end{equation}
Here $P_{({\rm tr})} \equiv P_{(1)} = P_{(2)}$ and $P_{||} \equiv
P_{(3)}$ are the transversal and longitudinal pressure components,
respectively, coinciding with the corresponding eigenvalues of the
{\it material  part} of the total stress-energy tensor.

\subsection{Reduced Maxwell equations}

We assume that the solutions inherit the spacetime symmetry and
the quantities $F_{ik}$ and $M_{ik}$ are the function of
cosmological time only. Thus, it follows from the second subsystem
of Maxwell equations (\ref{maxwellR}) that the spatial components
of the Maxwell tensor are constant, i.e., $ F_{\alpha \beta} =
const $. The only non-vanishing term is $F_{12} = const$, since
the magnetic field points along the $x^3$ axis. Note that the
tetrad component of the magnetic field, $B(t)$, is connected with
the constant $F_{12}$ by the following relationship
\begin{equation}
B(t) \equiv F^{*}_{ik} U^k X^i = \frac{F_{12}}{a^2(t)} \equiv B_0
\left(\frac{a(t_0)}{a(t)}\right)^2 \,. \label{chi2}
\end{equation}
We assume, that the electric field, electric polarization and
magneto-electric cross-effect are absent and the medium is locally
neutral. These requirements guarantee that the Maxwell equations
are satisfied identically. Thus, in the model under consideration
the evolution of the magnetization is governed only by the
constitutive equation (\ref{dM}).

\subsection{Reduced constitutive equations}

When $U^i=\delta^i_0$ in the metric (\ref{metric}) the
acceleration vector $DU^i$ vanishes, yielding $D \eta^{ikl} = D
(\epsilon^{ikls} U_s ) = 0$. The magnetization four-vector has
only one component:
\begin{equation}
M^i \equiv M^{*ik}U_k = - X^i M \,, \quad M^{ik}= - \eta^{ikl}M_l
= \eta^{ikl}X_l M \,. \label{tetrad2}
\end{equation}
Extended constitutive equations (\ref{dM}) reduce to one equation
of the relaxation type
\begin{equation}
\tau \dot{M}(t) + M = \left( \frac{1}{\mu_{||}} - 1 \right) B_0
\frac{a^2(t_0)}{a^2(t)} \,, \label{CEm}
\end{equation}
where the dot denotes derivative with respect to time.

\subsection{Reduced Einstein's equations}

The gravity field equations reduce to the following system
\begin{equation}
\left( \frac{\dot{a}}{a} \right)^2  + 2 \frac{\dot{a}}{a}
\frac{\dot{c}}{c} = \Lambda + \kappa ( W  + {\cal X}) \,,
\label{Ein11}
\end{equation}
\begin{equation}
\frac{\ddot{a}}{a} + \frac{\ddot{c}}{c} + \frac{\dot{a}}{a}
\frac{\dot{c}}{c} = \Lambda - \kappa ( P_{({\rm tr})} + {\cal X} )
\,, \label{Ein22}
\end{equation}
\begin{equation}
2 \frac{\ddot{a}}{a} + \left( \frac{\dot{a}}{a} \right)^2 =
\Lambda - \kappa ( P_{||} - {\cal X} ) \,. \label{Ein33}
\end{equation}
Differentiation of Einstein's equations leads
to the conservation law
\begin{equation}
\dot{W} + 2 \left( \frac{\dot{a}}{a} \right) \left[W + P_{({\rm
tr})}\right]+ \left(\frac{\dot{c}}{c} \right) \left[W +
P_{||}\right] + \dot{{\cal X}} +  4 \left( \frac{\dot{a}}{a}
\right) {\cal X} = 0 \,. \label{conserva}
\end{equation}
Note that ${\cal X}$ and its derivative enter the conservation law
with the multiplier, which does not depend on $c(t)$ (on $a(t)$
only). The function ${\cal X}(t)$ is now
\begin{equation}
{\cal X}(t) = \frac{1}{2} B(t)\left[ M(t) + B(t) \right] =
\frac{1}{2} B^2_0 \ \frac{a^4(t_0)}{ a^4(t)} \left[ 1 +
\frac{a^2(t)}{a^2(t_0)} \frac{M(t)}{B_0} \right] \,. \label{X}
\end{equation}
When the magnetization $M$ vanishes, the quantity ${\cal X}(t)$ is
non-negative. Nevertheless, when $M$ is non-vanishing, ${\cal
X}(t)$ can be negative during some time interval, or be equal to
zero. The total energy density ${\cal W} = W+{\cal X}$ is assumed
to remain positive.

\section{Magnetic relaxation with variable time parameter $\tau(t)$}

The relaxation equation (\ref{CEm}) and the quantity ${\cal
X}(t)$, the source of the gravitational field (\ref{X}), contain
the function $a(t)$ and do not contain $c(t)$.  This means that it
is reasonable to split the master equations into two subsystems,
dealing with the longitudinal ($P_{||}$, $B(t)$, $M(t)$) and
transversal ($P_{({\rm tr})}$) quantities, as well as, with
functions $c(t)$ and $a(t)$, describing the evolution of the
Universe in the longitudinal and transverse directions,
respectively.

Notice that the sign of the right-hand-side of the equation
(\ref{CEm}) depends on the sign of the expression $(1 / \mu_{||} -
1)$. According to the standard definitions, when $\mu_{||} > 1$ we
deal with the so-called paramagnetic medium; in this case the
expression in the parentheses in (\ref{CEm}) is negative. When $0<
\mu_{||} < 1$, the medium can be indicated as diamagnetic, this
case relates to the positive expression in the parentheses in
(\ref{CEm}). Finally, when $\mu_{||}
>> 1$, one can say that the medium is in a ferromagnetic phase. We
do not consider the cases with negative and vanishing $\mu_{||}$.
In principle, $\mu_{||}$ can be treated as a function of time.
This means that the sign of the expression $(1 / \mu_{||}-1)$ may
change with time. This problem itself is very interesting, but the
numerical calculations are needed to solve the corresponding
master equations. We focus here on the analytical solutions, and
do not consider the case of variable $\mu_{||}$.

Generally, the master equations of the EEM-model are
self-consistent. In order to find $M(t)$ from (\ref{CEm}) we have
to know $a(t)$. In order to find $a(t)$ from (\ref{Ein33}) we
should know ${\cal X}(t)$ which includes $M(t)$. To solve such a
problem one should introduce some extra ansatz. We prefer to start
from the equation (\ref{CEm}), and our ansatz concerns the
function $\tau(t)$. There are at least three ways to introduce the
relaxation time parameter $\tau(t)$. The first approach (which is
the simplest one) assumes that $\tau$ is constant. We consider the
model with $\tau(t) = \tau_0$ in Section 6. In the second approach
the relaxation time $\tau(t)$, the bulk viscosity coefficient
$\zeta(t)$, etc., are modeled as a power-law functions of the
energy density scalar $W$ (see, e.g., \cite{Pavon91} -
\cite{Maartens95}). We do not consider such options in this paper.
In the third approach $\tau(t)$ is assumed proportional to the
inverse rate of expansion. For instance, in the isotropic
Friedmann model ($a(t)=c(t)$) the new dimensionless variable $\xi
= \tau H(t)$ is frequently used, where $H(t)= \dot{a} / a$ is the
Hubble expansion parameter (see, e.g., \cite{Z99}). In this
section we consider the third version of the representation of
$\tau(t)$.

Since only the function $a(t)$ enters the relaxation equation and
the function ${\cal X}(t)$, we consider $\tau(t)= \xi
\frac{a}{\dot{a}}$. It is convenient to use a new variable $x
\equiv \frac{a(t)}{a(t_0)}$ and a function $H_{(a)}(x) =
\frac{\dot{x}}{x}$, for which
\begin{equation}
\frac{\dot{a}}{a} = H_{(a)}(x) \,, \quad \frac{\ddot{a}}{a} =
\frac{1}{2} x \frac{d}{dx} H^2_{(a)} + H^2_{(a)} \,.
\label{xH}
\end{equation}
The three-parameter family of solutions to (\ref{CEm}) reads
\begin{equation}
M\left(x,\xi,M(t_0),\mu_{||}\right) = M_0 x^{- \frac{1}{\xi}} +
\frac{B_0}{(1-2\xi)} \left(\frac{1}{\mu_{||}} - 1 \right)
\left[x^{-2} - x^{-\frac{1}{\xi}}\right]
\,.
\label{1M1}
\end{equation}
Here $2\xi \neq 1$, the special case $\xi = \frac{1}{2}$ will be
considered in Subsection 4.2. The corresponding expression for
${\cal X}\left(x,\xi,M_0,\mu_{||}\right)$ is
\begin{equation}
{\cal X}\left(x,\xi,M_0,\mu_{||}\right) = \frac{1}{2} B^2_0 \left[
K_1\left(\xi,\mu_{||}\right) \ x^{-4} +
K_2\left(\xi,M_0,\mu_{||}\right) \ x^{-
\left(2+\frac{1}{\xi}\right)} \right] \,, \label{1X1}
\end{equation}
where
\begin{equation}
K_1\left(\xi,\mu_{||}\right) \equiv \frac{(1 - 2 \xi \mu_{||})}{\mu_{||}(1 - 2 \xi )}  \ \ \
{\rm and} \qquad
K_2\left(\xi,M_0,\mu_{||}\right) \equiv \frac{M_0}{B_0} +
\frac{(\mu_{||} - 1)}{\mu_{||}(1 - 2 \xi )}
\,,
\label{1K1K2}
\end{equation}
are constant. We use for simplicity the definitions $M_0 \equiv
M(t_0)$ and $B_0 \equiv B(t_0)$, and below we omit the parameters
$\xi$, $M_0$ and $\mu_{||}$ in the arguments of $M$, ${\cal X}$
and $K_1, K_2$.

\subsection{Non-resonant magnetic relaxation: $\xi > 0$, $\xi \neq \frac{1}{2}$ }

For such $\xi$ the magnetization decreases when the Universe expands, i.e.,
$M(t \to \infty) \to 0$ when $a(t)$ increases. The relaxation of the initial magnetization
$M_0$ is characterised by the function $x^{- 1 / \xi}$.
The relaxation of the induced magnetization $F_{12} (1 / \mu_{||} - 1)$ is characterised in
the limit
$x \to \infty$ by the function $x^{- 1 / \xi}$, when $\xi > \frac{1}{2}$, and by the function
$x^{- 2}$, when $\xi < \frac{1}{2}$.

\subsubsection{Extremums and zeros of the function $X(t)$}

The formula (\ref{1X1}) shows, that the behaviour of ${\cal X}(x)$
is monotonic when $K_1 K_2 >0$. When $K_1 K_2 <0$, there exists
one extremum at $t=t_{*}$ such that
\begin{equation}
\left(\frac{a(t_{*})}{a(t_0)} \right)^{2 {-} \frac{1}{\xi}} {=} {-} \frac{4 \xi K_1}{K_2 (2\xi +1)}
\,,
\quad X(t_{*}) {=}  \frac{B^2(t_{*})(1 - 2 \xi \mu_{||})}{2 \mu_{||} (2\xi +1)}
\,. \label{1extremum}
\end{equation}
In addition, the inequality $t_{*} > t_0$ is valid for $4\xi
\left|K_1 \right| > (2\xi +1) \left|K_2 \right|$. It is a
requirement to the ratio $M_0 / B_0$, and we suppose it is valid.
The extremum is a minimum with negative value ${\cal X}(t_{*})$,
when $ \frac{1}{\xi} < 2 \mu_{||}$ and a maximum with positive
value ${\cal X}(t_{*})$, when $\frac{1}{\xi} > 2 \mu_{||}$, for
both paramagnetic and diamagnetic medium. Taking into account the
definitions of $K_1$ and $K_2$, (\ref{1K1K2}), one can find the
following possibilities for ${\cal X}(t)$ to have an extremum.
When the medium is paramagnetic, i.e., $\mu_{||}>1$, one obtains
three different situations.

\noindent
({\it i}) $\frac{1}{\xi} < 2$, $K_2 <0$.

\noindent In this case there is a minimum. The curve ${\cal X}(t)$
passes its zero-value point, if ${\cal X}(t_0) >0$, and tends
asymptotically to zero as ${\cal X}(t) \propto a^{-2 -
\frac{1}{\xi}}$, when $a(t) \to \infty $.

\noindent
({\it ii}) $\frac{1}{\xi} > 2\mu_{||}$, $K_2 <0$.

\noindent This case corresponds to a maximum. The curve ${\cal
X}(t)$ passes its zero-value point, if ${\cal X}(t_0) < 0$, and
tends asymptotically to zero as ${\cal X}(t) \propto a^{-4}$, when
$a(t) \to \infty $.

\noindent
({\it iii}) $2 < \frac{1}{\xi} < 2\mu_{||}$, $K_2 > 0$.

\noindent This case is analogous to the first one.

\noindent
Likewise, for the diamagnetic medium $\mu_{||}<1$ one obtains the following situations.

\noindent
({\it i}) $\frac{1}{\xi} > 2$, $K_2 <0$ (maximum).

\noindent
({\it ii}) $\frac{1}{\xi} < 2\mu_{||}$, $K_2 <0$ (minimum).

\noindent
({\it iii}) $2 > \frac{1}{\xi} > 2\mu_{||}$, $K_2 > 0$ (maximum).

\noindent Besides, there exists a time moment, $t_{**}$, when the
source ${\cal X}$ vanishes. According to (\ref{1X1}) the condition
${\cal X}(t_{**})=0$ can be satisfied, when
\begin{equation}
\left(\frac{a(t_{**})}{a(t_0)} \right)^{2 {-} \frac{1}{\xi}} {=}
{-} \frac{K_1}{K_2} \,, \label{1zero}
\end{equation}
but this is possible if the constants $K_1$ and $K_2$ have
opposite signs. Thus, the existence of zeros of ${\cal X}(t)$
assumes the same requirements $K_1 K_2 < 0$, as conditions for the
existence of extremums.

\subsubsection{Asymptotic case: fast relaxation, $\xi \to 0$}

When $\xi \to 0$ the formulas  (\ref{1M1}) and (\ref{1X1}) give
\begin{equation}
M(x)_{|\xi \to 0} = B_0 \ x^{-2} \left(\frac{1}{\mu_{||}} - 1
\right) \,, \quad {\cal X}(x)_{|\xi \to 0} =
\frac{B^2(t_0)}{2\mu_{||}} x^{-4} \,. \label{1MXfast}
\end{equation}
In a paramagnetic medium  the ratio $M(x)_{|\xi \to 0} / B_0$ is
negative, in a diamagnetic medium it is positive. In both cases $
{\cal X}(x)_{|\xi \to 0}$ is positive.

\subsubsection{Asymptotic case: slow relaxation, $\xi \to \infty$}

\noindent
When $\xi \to \infty$, one obtains
\begin{equation}
M(x)_{|\xi \to \infty} = M_0 \,, \quad {\cal X}(x)_{|\xi \to
\infty} = \frac{1}{2} B_0 \ x^{-4} \left( B_0 + M_0 x^{2} \right)
\,. \label{1MXslow}
\end{equation}
In these formulas $\mu_{||}$ is absent.

\subsubsection{Particular case: $2 \xi \mu_{||}=1$}

Studying the extremums of the function ${\cal X}(t)$ we have found
that the condition $2 \xi \mu_{||}=1$ (or, equivalently, $K_1 =
0$) gives some threshold value for the parameter $\xi$. Let us
consider this particular case separately. When $2 \xi \mu_{||}=1$
\begin{equation}
M(x)_{|\xi = \frac{1}{2 \mu_{||}}} = (M_0+B_0) x^{- 2 \mu_{||}} - B_0  x^{-2}
\,,
\label{1M123}
\end{equation}
and the function ${\cal X}(t)$ monotonically tends to zero  as $x
\to \infty$
\begin{equation}
{\cal X}(x)_{|\xi = \frac{1}{2 \mu_{||}}} = \frac{1}{2} B_0 (B_0 +
M_0) \ x^{- 2\left(1+ \mu_{||}\right)} \equiv {\cal X}_0  \ x^{-
2\left(1+ \mu_{||}\right) } \,, \label{2ximu1}
\end{equation}
remaining positive or negative depending on the  sign of its
initial value ${\cal X}_0$. Obviously, the value ${\cal
X}(x)_{|\xi = \frac{1}{2} \mu_{||}}$ is positive or negative
depending on the sign of the sum $M_0 + B_0$, and is equal to zero
identically when $M(t_0) = - B(t_0)$. In the last case the
information about magnetic field and magnetization does not enter
the Einstein field equations and we deal with the so-called
``hidden" magnetic field \cite{BZ05}. Notice that in the
ferromagnetic phase $\mu_{||} >> 1$, which corresponds here to the
model of fast relaxation $\xi << 1$, the source ${\cal X}(t)$
disappears very quickly.

\subsubsection{Particular case: vanishing initial magnetization, $M_0 = 0$}

This particular case can be studied in more detail. The minimum of
the function ${\cal X}(t)$ exists at the point given by
\begin{equation}
\left(\frac{a(t_{*})}{a(t_0)} \right)^{2 -\frac{1}{\xi}} =
\frac{4 \xi (2 \xi \mu_{||} - 1)}{(2\xi +1)(\mu_{||} - 1)}
\,.
\label{1M0zero}
\end{equation}
Since we put $M_0=0$, we should check especially the condition
$a(t_{*}) > a(t_0)$. When the medium is paramagnetic, this
condition is satisfied for $\xi > \frac{1}{2}$. This corresponds
to the first case of the classification of the extremums: the
function ${\cal X}(t)$ starts from  the positive initial value
${\cal X}_0$, takes its zero value at $t_{**}$, reaches its
minimum at $t_{*}$ and tends to zero asymptotically, remaining
negative. Here
\begin{equation}
{\cal X}_0 = \frac{B^2_{0}}{2} \,, \quad
\left(\frac{a(t_{**})}{a(t_0)} \right)^{2 -\frac{1}{\xi}} =
\frac{(2 \xi \mu_{||} - 1)}{(\mu_{||} - 1)} > 1 \,,
\label{1XMOzero}
\end{equation}
\begin{equation}
{\cal X}(t_{*}) = \frac{B^2_{0} (1 - 2 \xi \mu_{||})}{2 \mu_{||}
(2\xi +1)} \left[\frac{4 \xi (2 \xi \mu_{||} - 1)}{(\mu_{||} -
1)(2 \xi + 1)} \right]^{\frac{4\xi}{1-2\xi}} < 0 \,.
\label{1XextremumM0zero}
\end{equation}
The inequalities for the diamagnetic medium can be obtained analogously.

\subsection{Resonance magnetic relaxation: $\xi = \frac{1}{2}$ }

When $\xi = \frac{1}{2}$, the special solution to (\ref{CEm})
exists
\begin{equation}
M(x)_{|\xi = \frac{1}{2}} = x^{-2} \left[ M_0 +
2 B_0 \left(\frac{1}{\mu_{||}} - 1 \right) \log{x} \right]
\,,
\label{1Mlog}
\end{equation}
which tends to zero at $t \to \infty$ ($x \to \infty$). The term
"resonance" relates to special case when the relaxation parameter
$\tau(t)$ coincides with half of the characteristic expansion time
$1 / H_{(a)}$, where $H_{(a)} \equiv \frac{\dot{a}}{a}$. The
function ${\cal X}(x)_{| \xi = \frac{1}{2}}$ reads
\begin{equation}
{\cal X}(x)_{|\xi = \frac{1}{2} } = \frac{1}{2} B_0 \ x^{-4}
\left[ B_0 + M_0 + 2 B_0 \left(\frac{1}{\mu_{||}} - 1 \right)
\log{x} \right] \,. \label{1Xlog}
\end{equation}
The function ${\cal X}(x)_{|\xi = \frac{1}{2}}$ reaches its
extremum value
\begin{equation}
{\cal X}(t_{*})= \frac{B^2(t_{*})}{4} \left(\frac{1}{\mu_{||}} - 1
\right) \,, \label{1Xlogminimal}
\end{equation}
when
\begin{equation}
\log\left(\frac{a(t_{*})}{a(t_0)} \right) = \frac{1}{4} +
 \frac{\mu_{||}}{2\left( \mu_{||} - 1 \right)} \left( 1 + \frac{M_0}{B_0} \right)
\,.
\label{1Xlogtstar}
\end{equation}
In a paramagnetic medium this extremum is the minimum,
the condition $a(t_{*}) > a(t_0)$ assumes that $M_0 / B_0 > (1 - 3 \mu_{||}) / 2 \mu_{||}$.
When the medium is diamagnetic, the extremum is the maximum,
the condition $a(t_{*}) > a(t_0)$ is satisfied if  $M_0 / B_0 < (1 - 3 \mu_{||}) / 2 \mu_{||}$.

\subsection{ Magnetic instability: $\xi < 0$ and
$M_0 \neq \frac{B_0 (1-\mu_{||})}{\mu_{||}(1+2|\xi|)}$ }

In this case the magnetization
\begin{equation}
M(x)_{|\xi<0} = M_0 x^{\frac{1}{|\xi|}} +
\frac{B_0}{(1+2|\xi|)} \left(\frac{1}{\mu_{||}} - 1 \right)
\left[x^{-2} - x^{\frac{1}{|\xi|}}\right]
\,.
\label{1M111}
\end{equation}
increases as $x^{ 1 / |\xi|}$. In the special case
$M_0 = \frac{B_0 (1-\mu_{||})}{\mu_{||}(1+2|\xi|)}$ the term $x^{\frac{1}{|\xi|}}$
disappears, and the magnetization decreases as $x^{-2}$.
The function
\begin{equation}
{\cal X}(x)_{|\xi<0} = \frac{1}{2} B^2_0 \left[ K_1 \ x^{-4} + K_2
\ x^{-2+\frac{1}{|\xi|}} \right] \label{1Xinstab}
\end{equation}
decreases when $\frac{1}{|\xi|} < 2$ and increases when $\frac{1}{|\xi|}>2$.
When $\frac{1}{|\xi|} = 2$ it tends asymptotically to the constant value
\begin{equation}
{\cal X}\left(\infty \right) = \frac{1}{2} B_0 \left[ M_0 +  B_0
\frac{(\mu_{||} - 1)}{2\mu_{||}} \right] \,. \label{1Xinftyinstab}
\end{equation}

\section{Cosmological evolution in case of variable relaxation time}

Below we discuss exactly solvable models of four selected types,
describing cosmological evolution of the magnetizable medium. In
order to explain our choice let us note that when ${\cal X}$
depends on time via $a(t)$ only, the equation (\ref{Ein33}) does
not contain the information about $c(t)$, if $P_{||}$ also depends
on $a$ only, i.e., $P_{||}=P_{||}(a(t))$. The vacuum-type equation
of state $W + P_{||}=0$ belongs to this class of models, since
$c(t)$ disappears from conservation law (\ref{conserva}), and its
solution of the form $W(a)=-P_{||}(a)$ pics out (\ref{Ein33}) from
other Einstein's field equations.

\subsection{First example of cosmological evolution: \\ paramagnetic / diamagnetic dust}

When the medium behaves as a dust, i.e., $P_{||}= P_{({\rm tr})} =
0$, the optimal way to solve the master equations is the
following. First, we solve (\ref{Ein33}), which in terms of
variable $x$ takes the form of the equation for $H^2_{(a)}(x)$
\begin{equation}
x^{-2} \frac{d}{dx} \left( x^3 H^2_{(a)} \right)  = \Lambda + \frac{1}{2} \kappa B^2_0
\left[ K_1 \ x^{-4} + K_2 \ x^{- \left(2+\frac{1}{\xi}\right)}
\right] \,.
\label{1xH}
\end{equation}
Second, we search for the function $a(t)$ using the quadratures
\begin{equation}
t-t_0 =
\pm \int^{\frac{a(t)}{a(t_0)}}_1 \frac{dx}{x H_{(a)}(x)} \,.
\label{11t}
\end{equation}
When $a(t)$ is found, it is convenient to search for $c(t)$ using the substitution
$Y(t) = \frac{c(t)}{c(t_0)} \sqrt{\frac{a(t)}{a(t_0)}}$.
The equation governing the evolution of $Y(t)$ follows from (\ref{Ein22}) and (\ref{Ein33})
\begin{equation}
\ddot{Y}(t) + Y(t) \left[ - \frac{3}{4} \Lambda + \frac{5}{4} \kappa X(a(t)) \right] = 0 \,,
\label{c}
\end{equation}
and is a linear differential equation of the second order with
coefficient depending on time. The initial data for $Y(t)$ follow
from the definition and from the equation (\ref{Ein11})
\begin{equation}
Y(t_0) = 1 \,, \quad \dot{Y}(t_0) = \frac{\Lambda + \kappa (W_0 +
{\cal X}_0)}{2 H_{(a)}(t_0)} \,. \label{init}
\end{equation}
Finally, we search for $W(t)$ using (\ref{Ein11}) with obtained $a(t)$, $c(t)$, $\dot{a}$ and
$\dot{c}$.
In the process of integration of the equation (\ref{1xH}) a new resonance value of the parameter
$\xi$, namely $\xi = 1$, appears.  Indeed, when $\xi \neq 1$,  the solution is
\begin{equation}
H_{(a)}(x) = \pm \sqrt{\frac{\Lambda}{3}  + K_3 \ x^{-3} -
\frac{\kappa B^2_0}{2} \left[  K_1 \ x^{-4} + K_2 \frac{\xi}{(1 - \xi)} \
x^{- \left(2 + \frac{1}{\xi} \right)} \right] }
\,,
\label{111H}
\end{equation}
where
\begin{equation}
K_3 \equiv  H^2_{(a)}(t_0) -
\frac{\Lambda}{3} + \frac{\kappa B^2_0}{2} \left[ K_1 + K_2 \frac{\xi}{(1 - \xi)} \right]
\,.
\label{12H}
\end{equation}
When $\xi =1$, i.e., the relaxation time parameter $\tau(t)$ coincides with $H^{-1}_a$,
we should replace (\ref{111H}) by
\begin{equation}
H_{(a)}(x) = \pm \sqrt{\frac{\Lambda}{3}  + \tilde{K}_3 \ x^{-3} -
\frac{\kappa B^2_0}{2} \left[  K_1 \ x^{-4} + K_2 x^{-3} \log{x} \right] }
\,,
\label{11H1}
\end{equation}
where
\begin{equation}
\tilde{K}_3 \equiv  H^2_{(a)}(t_0) -
\frac{\Lambda}{3} + \frac{\kappa B^2_0}{2}  K_1
\,.
\label{12H1}
\end{equation}

\subsubsection{General properties of solution $H_{(a)}(t)$}

\vspace{3mm}
\noindent
{\it Asymptotics}

\noindent The quantity $H_{(a)}$ plays a role of Hubble function,
describing the Universe evolution in the plane $x^1Ox^2$. When
$\Lambda \neq 0$ and $\frac{1}{\xi}>-2$, the (positive) asymptotic
value of this function, $H_{a}(\infty) =
\sqrt{\frac{\Lambda}{3}}$, is well-known for the Friedmann
isotropic  model. The corresponding asymptotic behaviour of $a(t)$
is $a(t) \propto \exp\{\sqrt{\frac{\Lambda}{3}} \ t \}$. When
$\Lambda = 0$, the asymptotic formula for $H_{(a)}$ is
predetermined by the value of the parameter $\xi$:

\noindent
a) if $\frac{1}{\xi} < 1$, $H_{(a)} \propto x^{- \left(1+\frac{1}{2\xi} \right)}$ and
$a(t) \propto t^{ \frac{2\xi}{2\xi +1}}$

\noindent
b) if $\frac{1}{\xi} > 1$, $H_{(a)} \propto x^{- \frac{3}{2}}$, and
$a(t) \propto t^{\frac{2}{3}}$.

\noindent
When $\frac{1}{\xi}=-2$ the last term in the expression for
$H_{(a)}$ (\ref{111H}) does not depend on time and redefines
effectively the cosmological constant
\begin{equation}
\Lambda \to \Lambda^{*} = \Lambda + \frac{\kappa B^2_0}{2} \left( \frac{M_0}{B_0} +
\frac{\mu_{||}-1}{2\mu_{||}} \right) \,.
\label{cc}
\end{equation}

\noindent When $\frac{1}{\xi}<-2$ the last term in the expression
for $H_{(a)}$ (\ref{111H}) is the leading order term at $t \to
\infty$, and the Universe collapses in the cross-section
$x^1Ox^2$. It is an exotic case, and we do not focus on it.

\vspace{3mm}
\noindent
{\it Extremums}

\noindent Generally, the behaviour of the $H_{(a)}(x)$ function is
non-monotonic, i.e., there are intervals with $\dot{H}_{(a)}(t) <
0$ as well as with $\dot{H}_{(a)}(t) > 0$. The necessary condition
for the existence of extremums marks the points $x_{(1)},
x_{(2)},... x_{(s)}$, in which
\begin{equation}
K_3 \ x_{(s)} -
\frac{\kappa B^2_0}{6} \left[ 4 K_1  {+} K_2 \frac{2\xi{+}1}{(1 {-} \xi)}
x_{(s)}^{ \left(2 {-} \frac{1}{\xi} \right)} \right] = 0
\,.
\label{11H}
\end{equation}
Formally speaking, this equation can give $(m)$ real roots, the
function $H_{(a)}$ can possess $(m)$ extremums, and, consequently,
$(m)$ points can appear, in which the transition from the
accelerated expansion to the decelerated one (and vice-versa)
takes place. To illustrate these possibilities, consider now
several particular cases.

\subsubsection{$K_1 = K_3 = 0$}

Let us choose the initial values
$H_{(a)}(t_0)$, $M_0$, $F_{12}$, $a(t_0)$, as well as, $\xi$ and $\mu_{||}$ parameters in an
appropriate manner to provide the relations $K_1 = 0$ and $K_3 = 0 $. It is possible when
\begin{equation}
\frac{1}{\xi} = 2 \mu_{||} \,, \quad
\frac{\Lambda}{3} =
H^2_{(a)}(t_0) + \frac{\kappa B_0 (B_0 + M_0)}{2 (2 \mu_{||} - 1)} \,,
\quad K_2 = \frac{B_0 + M_0}{B_0}
\,.
\label{1simpl}
\end{equation}
For this case ${\cal X}(x)$ is the monotonic function
\begin{equation}
{\cal X}(t) = \frac{1}{\kappa}\left( 1- 2 \mu_{||} \right)
\left(H^2_{(a)}(t_0) - \frac{\Lambda}{3}\right) x^{ - 2(1 +
\mu_{||})}  \,, \label{11Xsimpl}
\end{equation}
and $H_{(a)}(x)$ simplifies essentially
\begin{equation}
H_{(a)}(x) = \pm \sqrt{
\frac{\Lambda}{3} + \left(H^2_{(a)}(t_0) - \frac{\Lambda}{3}\right)
x^{ - 2(1 + \mu_{||})} } \,.
\label{11Hmono}
\end{equation}
Here $\mu_{||} \neq \frac{1}{2}$ to avoid the relation $\xi = 1$.
For such $H_{(a)}(x)$ the equation (\ref{11t}) can be easily integrated
\begin{equation}
\frac{a(t)}{a(t_0)} {=}  \left\{\cosh {\left[\sqrt{\frac{\Lambda}{3}}(1{+}\mu_{||})(t{-}t_0)\right]} +
H_{(a)}(t_0) \sqrt{\frac{3}{\Lambda}}
\sinh \left[\sqrt{\frac{\Lambda}{3}}(1{+}\mu_{||})(t{-}t_0)\right]
\right\}^{\frac{1}{1{+}\mu_{||}}} \,.
\label{11atu}
\end{equation}
$H_{(a)}(t)$ takes an explicit form
\begin{equation}
H_{(a)}(t) =   \sqrt{\frac{\Lambda}{3}} \left\{
\frac{ H_{(a)}(t_0) +  \sqrt{\frac{\Lambda}{3}}
\tanh{\left[\sqrt{\frac{\Lambda}{3}}(1+\mu_{||})(t-t_0)\right]}}{ \sqrt{\frac{\Lambda}{3}} +
H_{(a)}(t_0) \tanh{\left[\sqrt{\frac{\Lambda}{3}}(1+\mu_{||})(t-t_0)\right]} } \right\}
\,.
\label{11Htu}
\end{equation}
%\noindent
The asymptotic behaviour of this solution is given by
\begin{equation}
a(t \to \infty) \Rightarrow  a(t_0)
\left[\frac{1}{2}\left( 1 + H_{(a)}(t_0)
\sqrt{\frac{3}{\Lambda}} \right)\right]^{\frac{1}{1+\mu_{||}}}
\exp{\sqrt{\frac{\Lambda}{3}}(t-t_0)}
 \,,
\label{11aasympt}
\end{equation}
\begin{equation}
H_{(a)}(t \to \infty) \rightarrow \sqrt{\frac{\Lambda}{3}}  \,.
\label{11Hasympt}
\end{equation}
Returning to the function $c(t)$, note that the equation (\ref{c})
with $a(t)$, given by (\ref{11atu}), is the Hill equation
\cite{Stoker} with imaginary argument. Since ${\cal X}(t \to
\infty) \to 0$, the appropriate asymptotics of $Y(t)$ and $c(t)$
are
\begin{equation}
Y(t \to \infty) \propto \exp\left\{ \frac{3}{2} \sqrt{\frac{\Lambda}{3}} \ t \right\} \,,
\quad c(t \to \infty) \propto \exp\left\{ \sqrt{\frac{\Lambda}{3}}  \ t \right\}
\,,
\label{Yasympt}
\end{equation}
thus, the Universe isotropizes.
To specify $c(t)$ consider three special examples.

\vspace{3mm}
\noindent
{\it First special example: $H_{(a)}(t_0) \equiv \sqrt{\frac{\Lambda}{3}}$}

\noindent When $H_{(a)}(t_0) \equiv \sqrt{\frac{\Lambda}{3}}$,
${\cal X}=0$, $M_0 + B_0 =0$ and the behaviour of $a(t)$ is
governed by the de Sitter expansion law
\begin{equation}
a(t) = a(t_0) \exp\left\{\sqrt{\frac{\Lambda}{3}}(t-t_0)\right\}
\,.
\label{adesitter}
\end{equation}
The solution for $c(t)$ is
\begin{equation}
c(t) = \left[ c(t_0) + \frac{\kappa W_0}{2 \Lambda a^2(t_0)} \right]
\exp\left\{\sqrt{\frac{\Lambda}{3}}(t-t_0)\right\}
- \frac{\kappa W_0}{2 \Lambda a^2(t_0)}
\exp\left\{ -2 \sqrt{\frac{\Lambda}{3}}(t-t_0)\right\}
\,.
\label{cdesitter}
\end{equation}
The asymptotic behaviour of $c(t)$ at $t \to \infty$ is the same as for $a(t)$.
This solution corresponds to the model of hidden induction, mentioned in the subsubsection
4.1.4. The energy density scalar behaves as $W(t) = W_0 \frac{c(t_0) a^2(t_0)}{c(t) a^2(t)}$.

\vspace{3mm}
\noindent
{\it Second special example: $H_{(a)}(t_0) = 0$}

\noindent
When $H_{a}(t_0) \equiv 0$, one obtains the simplified formulas
\begin{equation}
H_{(a)}(t) =   \sqrt{\frac{\Lambda}{3}}
 \ \ \tanh{\left[\sqrt{\frac{\Lambda}{3}}(1+\mu_{||})(t-t_0)\right]} \,,
\label{11Hzeroinitial}
\end{equation}
\begin{equation}
\frac{a(t)}{a(t_0)} =
\left[ \cosh{\sqrt{\frac{\Lambda}{3}}(1+\mu_{||})(t-t_0)} \right]^{\frac{1}{1 + \mu_{||}}}
\,,
\label{12at}
\end{equation}
\begin{equation}
\kappa {\cal X}(t) = \frac{\Lambda}{3} \left( 2 \mu_{||} -1
\right)
\cosh^{-2}\left[{\sqrt{\frac{\Lambda}{3}}(1+\mu_{||})(t-t_0)}
\right]  \,. \label{123Xsimpl}
\end{equation}
The equation for $Y$ can be transformed into the Legendre
equation
\begin{equation}
(1-z^2)Y''(z) - 2 z Y'(z) + Y \left[ \nu (\nu +1) - \frac{\lambda^2}{1-z^2} \right] = 0 \,,
\label{legendre}
\end{equation}
where
\begin{equation}
z = \tanh\left[{\sqrt{\frac{\Lambda}{3}}(1{+}\mu_{||})(t{-}t_0)} \right] \,, \quad
\nu (\nu {+}1) \equiv \frac{5 (2 \mu_{||} {-} 1)}{4 (1{+}\mu_{||})^2} \,, \quad
\lambda^2 = \frac{9}{4(1{+}\mu_{||})^2} \,.
\label{12Xsimpl}
\end{equation}
Thus, one obtains
\begin{equation}
\frac{c(t)}{c(t_0)} =
\left[ 1- z^2(t)\right]^{ \frac{1}{4(1 + \mu_{||})}}
\left\{
{\cal C}_1 {\cal P}^{\lambda}_{\nu}(z(t)) +  {\cal C}_2 {\cal Q}^{\lambda}_{\nu}(z(t))
\right\}
\,,
\label{cLegendre}
\end{equation}
where
${\cal P}^{\lambda}_{\nu}(z)$ and ${\cal Q}^{\lambda}_{\nu}(z)$ are the associated Legendre
functions of the first and second kinds, respectively (see, \cite{AbSt}, 8.1.1).
${\cal C}_1$ and ${\cal C}_2$ are the constants of integration
\begin{equation}
{\cal C}_1 = \frac{1}{Wr(0)}
\left[
({\cal Q}^{\lambda}_{\nu})'(0) - J {\cal Q}^{\lambda}_{\nu}(0)
\right] \,, \quad
{\cal C}_2 = \frac{1}{Wr(0)}
\left[
J {\cal P}^{\lambda}_{\nu}(0) - ({\cal P}^{\lambda}_{\nu})'(0)
\right] \,,
\label{cLegendre1}
\end{equation}
where
\begin{equation}
Wr(0) \equiv
{\cal P}^{\lambda}_{\nu}(0) ({\cal Q}^{\lambda}_{\nu})'(0) {-}
({\cal P}^{\lambda}_{\nu})'(0) {\cal Q}^{\lambda}_{\nu}(0)
{=} \frac{2^{2\lambda} \Gamma(\frac{1}{2}\nu + \frac{1}{2}\lambda +1)
\Gamma(\frac{1}{2}\nu + \frac{1}{2}\lambda +\frac{1}{2})}{
\Gamma(\frac{1}{2}\nu - \frac{1}{2}\lambda + 1)\Gamma(\frac{1}{2}\nu - \frac{1}{2}\lambda +
\frac{1}{2})}
\,,
\label{cLegendre11}
\end{equation}
is the Wronsky determinant at $z=0$ ($t=t_0$), $\Gamma(q)$ is Gamma-function,
\begin{equation}
J = \frac{\dot{c}(t_0)}{c(t_0)} \sqrt{\frac{3}{\Lambda}} (1+\mu_{||})^{-1}\,,
\label{cLegendre12}
\end{equation}
and
$$
{\cal P}^{\lambda}_{\nu}(0) =
2^{\lambda} \pi^{-\frac{1}{2}}
\frac{\Gamma(\frac{1}{2}\nu + \frac{1}{2}\lambda +\frac{1}{2})}{\Gamma(\frac{1}{2}\nu
- \frac{1}{2}\lambda + 1)} \ \cos{\left[\frac{\pi (\nu + \lambda)}{2}\right]} \,,
$$
$$
({\cal P}^{\lambda}_{\nu})'(0) =
2^{\lambda+1} \pi^{-\frac{1}{2}}
\frac{\Gamma(\frac{1}{2}\nu + \frac{1}{2}\lambda + 1)}{\Gamma(\frac{1}{2}\nu
- \frac{1}{2}\lambda + \frac{1}{2})} \ \sin{\left[\frac{\pi (\nu + \lambda)}{2}\right]} \,,
$$
$$
{\cal Q}^{\lambda}_{\nu}(0) =
- 2^{\lambda-1} \pi^{\frac{1}{2}}
\frac{\Gamma(\frac{1}{2}\nu + \frac{1}{2}\lambda + \frac{1}{2})}{\Gamma(\frac{1}{2}\nu
- \frac{1}{2}\lambda + 1)} \ \sin{\left[\frac{\pi (\nu + \lambda)}{2}\right]}
\,,
$$
\begin{equation}
({\cal Q}^{\lambda}_{\nu})'(0) =
2^{\lambda} \pi^{\frac{1}{2}}
\frac{\Gamma(\frac{1}{2}\nu + \frac{1}{2}\lambda + 1)}{\Gamma(\frac{1}{2}\nu
- \frac{1}{2}\lambda + \frac{1}{2})} \ \cos{\left[\frac{\pi (\nu + \lambda)}{2}\right]}
\,.
\label{cLegendre13}
\end{equation}

\vspace{3mm}
\noindent
{\it Third special example: $ \Lambda = 0$}

\noindent
The solution of the problem is
\begin{equation}
\kappa {\cal X}(t) = \left( 1- 2 \mu_{||} \right) H^2_{(a)}(t_0)
 \ x^{ - 2(1 + \mu_{||})}  \,, \quad
H_{(a)}(x) = \pm H_{(a)}(t_0) \ x^{ - (1 + \mu_{||})}  \,,
\label{11Hmonotonic}
\end{equation}
\begin{equation}
\frac{a(t)}{a(t_0)} =  z^{\frac{1}{1+\mu_{||}}}(t) \,, \quad
H_{(a)}(t) =   \frac{ H_{(a)}(t_0)}{ z(t)} \,, \quad
z(t) \equiv 1 + H_{(a)}(t_0)(1+\mu_{||})(t-t_0) \,.
\label{11at}
\end{equation}
The equation (\ref{c}) reduces to the Euler equation
\begin{equation}
z^2 Y''(z) + Y \ \frac{5 (1 - 2 \mu_{||})}{4 (1 + \mu_{||})^2} = 0
\,,
\label{Euler}
\end{equation}
and $c(t)$ reads
\begin{equation}
c(t) = c(t_0) z^{ \frac{\mu_{||}}{2(1+\mu_{||})}} \left[
z^{- \sigma}  + {\cal C}_3 \left( z^{\sigma} - z^{- \sigma} \right) \right]\,,
\label{11Ht}
\end{equation}
where
\begin{equation}
\sigma = \frac{\sqrt{\mu^2_{||} + 12 \mu_{||} - 4}}{2(1+\mu_{||})} \,, \quad
{\cal C}_3 = \frac{1}{2} + \left[ \frac{\dot{c}(t_0)}{H_{(a)}(t_0) c(t_0)} -
\frac{\mu_{||}}{2}\right] \left(\mu^2_{||} + 12 \mu_{||} - 4\right)^{- \frac{1}{2}} \,.
\label{sigma12}
\end{equation}
This solution may not appear in the model without magnetization, since the necessary condition
$2H^2_{(a)}(t_0) (2\mu_{||}-1) + \kappa B_0 (M_0 + B_0) = 0$ is not valid when $M_0=0$ and
$\mu_{||} = 1$.
Note that the asymptotics for $a(t)$ and $c(t)$,
\begin{equation}
a(t \to \infty) \propto t^{\frac{1}{1+\mu_{||}}} \,, \quad
c(t \to \infty) \propto t^{\frac{\mu_{||} + \sqrt{\mu^2_{||} + 12 \mu_{||} - 4}}{2(1+\mu_{||})}} \,,
\label{sigma112}
\end{equation}
coincide only in the critical regime $\mu_{||} \to \frac{1}{2}$ or in other words, when
$\xi \to 1$. Thus, when $\Lambda =0$ the Universe does not isotropize for arbitrary
$\mu_{||} \neq \frac{1}{2}$.

\subsection{Second example of cosmological evolution: longitudinal quasi-vacuum}

When $W+P_{||}= 0$, the right-hand sides of (\ref{Ein11}) and of (\ref{Ein33}) coincide,
and their left-hand-sides give
\begin{equation}
\frac{\dot{c}}{c} \frac{\dot{a}}{a} = \frac{\ddot{a}}{a} \,.
\label{ccaa}
\end{equation}
One should distinguish two cases: first, when $\dot{a}(t) \neq 0$ and $H_{(a)}(t_0) \neq 0$, second,
when $\dot{a} = 0$.

\subsubsection{$a(t) = const = a(t_0)$}

The Einstein equations, supplemented by the equations of state
$P_{||} = - W$ and $P_{{\rm tr}} = (\gamma-1)W$, formally admit
the solution $a(t)=a(t_0)$, if, first, the quantities $P_{||}$,
$P_{{\rm tr}}$, $W$ and ${\cal X}$ are constant and are linked by
the relation $\Lambda + \kappa (W + {\cal X})) = 0$, second, the
equation for $c(t)$ has the form
\begin{equation}
\ddot{c} + c(t) Q = 0 \,, \quad Q \equiv \kappa (2-\lambda) {\cal
X}(a(t_0)) - \gamma \Lambda = const \,. \label{ddc}
\end{equation}
When $Q>0$, $c(t)$ oscillates harmonically with the frequency $\sqrt{Q}$, and this model is
singular. When $Q<0$, $c(t)$ behaves exponentially. At $Q=0$ $c(t)$ is a linear function of
time and  this model can be effectively reduced to the Minkowski spacetime.

\subsubsection{ $\dot{a} \neq 0$, $H_a(t_0) \neq 0$}

In this case the consequence (\ref{ccaa}) gives $c(t)$ readily in
terms of $a(t)$ and $H_{(a)}(t)$
\begin{equation}
c(t) = c(t_0) \frac{H_{(a)}(t)}{H_{(a)}(t_0)} \frac{a(t)}{a(t_0)} \,.
\label{caH}
\end{equation}
Asymptotic behaviour of $c(t)$ at $t \to \infty$ is the same as
for  $a(t)$ (i.e., the Universe isotropizes), when $H_{(a)}(t) \to
const$. The optimal strategy to obtain the solution is now the
following. First, we find $W$ from the conservation law
(\ref{conserva}) transformed into
\begin{equation}
\frac{d}{dx} \left( x^{2\gamma} W(x)\right) = - x^{2\gamma - 4}
\frac{d}{dx} \left( x^{4} {\cal X}(x)\right) \,. \label{dWW}
\end{equation}
Second, we solve the equation for $H^2_{a}(x)$
\begin{equation}
x^{-2} \frac{d}{dx} \left( x^3 H^2_{(a)} \right) = \Lambda +
\kappa [W(x) + {\cal X}(x)] \,, \label{HHH}
\end{equation}
which is the direct consequence of the equation  (\ref{Ein33}).
Third, we consider the solution of (\ref{11t}) for $a(t)$, and
then return to the solution (\ref{caH}) for $c(t)$.

When $2(\gamma-1) \xi \neq 1$, the solution of (\ref{dWW}) is
\begin{equation}
W(x) = W_0 x^{-2\gamma} + \frac{B^2_0 K_2 (2\xi-1)}{2[2\xi(\gamma-1)-1]}
\left[x^{-2\gamma} - x^{-\left(2 + \frac{1}{\xi} \right)} \right]
\,.
\label{WW}
\end{equation}
When  $2(\gamma-1) \xi = 1$, $W(x)$ behaves according to the formula
\begin{equation}
W(x) = x^{-\left(2 + \frac{1}{\xi} \right)} \left[
W_0 -
\frac{B^2_0 K_2 (2\xi-1)}{2\xi} \log{x} \right]
\,.
\label{WW1}
\end{equation}
In the first case, when $2(\gamma-1) \xi \neq 1$, the solution of (\ref{HHH}) takes the form
\begin{equation}
H^2_{(a)}(x) = \frac{\Lambda}{3} + x^{-3} L_1 + x^{-4} L_2 + x^{-2\gamma} L_3 +
x^{-2 - \frac{1}{\xi}} L_4
\,,
\label{HHH1}
\end{equation}
where
$$
L_1 = H^2_{(a)}(t_0) - \frac{\Lambda}{3} - L_2 - L_3 - L_4  \,,
$$
$$
L_2 = - \frac{1}{2} \kappa B^2_0 K_1 \,, \quad
L_3 = \frac{\kappa B^2_0}{2(3-2\gamma)} \left[ \frac{2 W_0}{B^2_0} +
K_2 \frac{(2\xi-1)}{[2\xi(\gamma-1)-1]} \right] \,,
$$
\begin{equation}
L_4 = \kappa B^2_0 K_2  \frac{\xi^2(\gamma-2)}{(\xi-1)[2\xi(\gamma-1)-1]}
\,.
\label{HHH2}
\end{equation}
Analogously to the case of paramagnetic / diamagnetic dust, we
attract attention to the fact that generally $H_{(a)}(x)$ is not a
monotonic function and has several extremums. This means that in
the expansion of the Universe there are time intervals
characterized by acceleration and deceleration. Asymptotic
behaviour of this function is predetermined by the parameters
$\Lambda$, $\gamma$ and $\xi$. The novelty in comparison with the
previous analysis is that a new parameter $\gamma$ is involved. To
obtain $a(t)$ we have to solve the equation (\ref{11t}) with
$H_{(a)}(x)$ given by (\ref{HHH1}). To illustrate our conclusions
let us consider two particular cases.

\subsubsection{Slow relaxation}

When $\xi \to \infty$ one obtains a model in which relaxation time
for the magnetization is much larger than the characteristic time
of the Universe evolution. Consider for simplicity $L_1=0$ and
$L_3=0$. It is possible if
\begin{equation}
M_0 = \frac{2 W_0 (1-\gamma)}{B_0} \,, \quad
H^2_{(a)}(t_0) = \frac{\Lambda}{3} + \kappa W_0 (2-\gamma) - \frac{1}{2} \kappa B^2_0
\,.
\label{L0L0}
\end{equation}
With such a choice of the initial parameters we have
\begin{equation}
H_{(a)}(x) = \sqrt{\frac{\Lambda}{3} + \kappa W_0 (2-\gamma) x^{-2} -
\frac{1}{2} \kappa B^2_0 x^{-4}}
\,,
\label{HdHu}
\end{equation}
\begin{equation}
H_{(a)}(x) \frac{d}{dx} H_{(a)}(x) = \kappa B^2_0 x^{-5} \left( 1 - \frac{x^2}{x^2_*} \right) \,.
\label{HdH1}
\end{equation}
Here the definition $x^2_* \equiv \frac{B^2_0}{W_0(2-\gamma)}$ is
used, and the assumptions $\gamma<2$ and $B^2_0 > W_0
(2{-}\gamma)$ are made. At the point $x{=}x_*$ the function
$H_{(a)}(x)$ reaches its maximum with
\begin{equation}
H_{(a)}(x_*) = \sqrt{\frac{\Lambda}{3} +  \frac{\kappa B^2_0}{2 x^{4}_*} }
\,,
\label{HdH}
\end{equation}
and tends to the value $\sqrt{\frac{\Lambda}{3}}$ asymptotically at $t \to \infty$.
Integration of (\ref{11t}) yields
$$
\left(\frac{a(t)}{a(t_0)} \right)^2 =
\cosh{\left[2\sqrt{\frac{\Lambda}{3}} \ (t{-}t_0) \right]} +
\sqrt{\frac{3}{\Lambda}} H_{(a)}(t_0)
\sinh{\left[2\sqrt{\frac{\Lambda}{3}} \ (t{-}t_0) \right]} +
$$
\begin{equation}
+  \frac{3 \kappa W_0 (2{-}\gamma)}{\Lambda}
\sinh^2{\left[\sqrt{\frac{\Lambda}{3}} \ (t{-}t_0) \right]} \,.
\label{xintegral}
\end{equation}
Thus, such a model gives $a(t)$, $H_{(a)}(t)$ and $c(t)$ in terms
of elementary (hyperbolic) functions and is convenient for
qualitative analysis.

\subsubsection{Transversal stiff matter, $\gamma=2$}

When $\gamma=2$, $P_{({\rm tr})} = W$, i.e., the matter behaves as
a stiff one. It follows from (\ref{dWW}) that
\begin{equation}
W(x) + {\cal X}(x) = x^{-4} (W_0 + {\cal X}_0) \,, \quad {\cal
X}_0 \equiv \frac{1}{2} B_0 (M_0 + B_0) \,, \label{WX}
\end{equation}
and (\ref{Ein33}) yields
\begin{equation}
H_{(a)}(x) = \sqrt{\frac{\Lambda}{3} + x^{-3} \left[H^2_{(a)}(t_0) - \frac{\Lambda}{3} +
\kappa (W_0 + X_0) \right]
- x^{-4} \kappa(W_0 + X_0)} \,.
\label{HHHH}
\end{equation}
This means that the parameter $\xi$ becomes hidden if we consider
the functions $H_{(a)}(t)$, $a(t)$ and $c(t)$.  Nevertheless, it
appears when we calculate $W(x)$:
\begin{equation}
W(x) = W_0 x^{-4} + \frac{1}{2} B^2_0 K_2 \left[ x^{-4} - x^{- \left(2+ \frac{1}{\xi} \right)}
\right] \,.
\label{WWWW}
\end{equation}
To illustrate the problem arising in a particular case
$H_{(a)}(t_0)=0$ mentioned in the beginning of Subsection 5.2,
consider now the solution (\ref{HHHH}) with a special choice of
initial data: $H_{(a)}(t_0)=0$ and $ \kappa(W_0 + {\cal X}_0) =
\frac{\Lambda}{3}$. Then one obtains
\begin{equation}
a(t) = a(t_0) \cosh^{\frac{1}{2}}{\left[2\sqrt{\frac{\Lambda}{3}} \ (t{-}t_0) \right]} \,, \quad
H_{(a)}(t) = \sqrt{\frac{\Lambda}{3}} \tanh{\left[2\sqrt{\frac{\Lambda}{3}} \ (t{-}t_0) \right]} \,,
\label{aH}
\end{equation}
and
\begin{equation}
c(t) = const \ \cosh^{-\frac{1}{2}}{\left[2\sqrt{\frac{\Lambda}{3}} \ (t{-}t_0) \right]}
\sinh{\left[2\sqrt{\frac{\Lambda}{3}} \ (t{-}t_0) \right]} \,.
\label{aHc}
\end{equation}
The model is self-consistent, if $c(t_0)=0$. The $const$ can be chosen from the isotropization
condition $a( t \to \infty) = c( t \to \infty)$, i.e., $const = a(t_0)$.

\section{Constant relaxation parameter}

\subsection{Evolution of magnetization}

When $\tau$ takes a constant value, $\tau_0$, we can readily solve (\ref{CEm}) in quadratures
to get
\begin{equation}
M(t) = \exp\left(- \frac{t-t_0}{\tau_0}\right) \left[ M(t_0) +
\frac{B_0 a^2(t_0)}{\tau_0} \left( \frac{1}{\mu_{||}} - 1 \right)
\int^t_{t_0} \frac{dt^{'}}{a^2(t^{'})} \exp\left(
\frac{t^{'}-t_0}{\tau_0}\right)\right] \,. \label{1M2}
\end{equation}
To obtain some analytical results consider the evolution equations
with $a(t)= a(t_0) e^{H_0(t-t_0)}$, where $H_0$ is a constant.
Such a two-dimensional de Sitter-type expansion is possible, e.g.,
when
\begin{equation}
P_{||}(t) = {\cal X}(t) \,, \quad  \Lambda =  3 H^2_0 \,.
\label{1Plong}
\end{equation}
In this case the total longitudinal pressure, ${\cal P}_{||}
\equiv P_{||}(t) - {\cal X}(t)$, vanishes. When $2H_0 \tau_0 \neq
1$ the formula (\ref{1M2}) for the magnetization and formula
(\ref{X}) for the ${\cal X}(t)$ read
$$
M(t) = \exp\left(- \frac{t-t_0}{\tau_0}\right) \left[ M(t_0) +
\frac{B_0 }{(1- 2H_0 \tau_0)} \left( 1 - \frac{1}{\mu_{||}} \right) \right]
$$
\begin{equation}
- \exp\left[- 2H_0 (t-t_0)\right]
\frac{B_0}{(1- 2H_0 \tau_0)} \left( 1 - \frac{1}{\mu_{||}} \right)
\,,
\label{12M}
\end{equation}
and
$$
{\cal X}(t) = \frac{B^2_{0}}{2} \left\{ \left[1 + \frac{1}{(2 H_0
\tau_0 -1)} \left( 1 - \frac{1}{\mu_{||}} \right) \right]
\exp{\left[- 4H_0(t-t_0) \right]} \right.
$$
\begin{equation}
\left.
+ \left[
\frac{M(t_0)}{B_{0}} -
\frac{1}{(2 H_0 \tau_0 -1)} \left( 1 - \frac{1}{\mu_{||}} \right)
\right]
\exp{\left[- \left(2H_0 + \frac{1}{\tau_0} \right)(t-t_0) \right]} \right\}
\,,
\label{12X}
\end{equation}
respectively. The asymptotic behaviour of ${\cal X}(t)$ is
dominated by the exponential function $\exp{\left[- 4H_0(t-t_0)
\right]}$ when $\frac{1}{\tau_0} > 2H_0$, (i.e., the double
relaxation time is less than the characteristic rate of expansion
$1/H_0$ ), and by another exponent $\exp{\left[- \left(2H_0 +
\frac{1}{\tau_0} \right)(t-t_0) \right]}$, when $\frac{1}{\tau_0}
< 2H_0$. Note that a special case with ${\cal X}(t) \equiv 0$
exists for
\begin{equation}
M(t_0) = - B_0 \,, \quad 2 H_0 \mu_{||} \tau_0 = 1
\,.
\label{12zeroX}
\end{equation}
The magnetization $M(t)$ for such a case decreases exponentially
\begin{equation}
M(t) = - B_0 \exp{[- 2H_0 (t-t_0)] }
\,.
\label{12Mspecial}
\end{equation}
In the resonance case, when $\frac{1}{\tau_0} = 2H_0$, the
formulas (\ref{1M2}) and (\ref{X}) give
\begin{equation}
M(t) = \exp\left(- \frac{t-t_0}{\tau_0}\right) \left[ M(t_0) -
\frac{B_{0} }{\tau_0} \left( 1 - \frac{1}{\mu_{||}} \right) (t-t_0) \right]
\,,
\label{12Mreson}
\end{equation}
\begin{equation}
{\cal X}(t) = \frac{B^2_{0}}{2} \exp{\left[-
\frac{2}{\tau_0}(t-t_0) \right]} \left[1 + \frac{M(t_0)}{B_{0}} +
\left( \frac{1}{\mu_{||}} - 1 \right)\left(\frac{t-t_0}{\tau_0}
\right)\right] \,. \label{12Xreson}
\end{equation}
Similarly to the case of time dependent relaxation parameter, the
function ${\cal X}(t)$ can be monotonic or non-monotonic,
depending on the value of the parameters $M(t_0)$, $H_0$,
$\tau_0$. To illustrate this fact consider the resonance case
$\frac{1}{\tau_0} = 2H_0$, when $\mu_{||}>1$. If the following
inequality takes place:
\begin{equation}
M(t_0) > \frac{B_{0}}{2 \mu_{||}} (1 - 3\mu_{||}) \,,
\label{12Millustra}
\end{equation}
the function ${\cal X}(t)$ decreases, passes through its zero
value, reaches a minimum at the point
\begin{equation}
t_{*}= t_0 + \frac{1}{4H_0 }
\left[ \frac{(3\mu_{||}-1)B_{0} + 2 \mu_{||} M(t_0) }{(\mu_{||}-1) B_{0}} \right]
\,,
\label{12tstar1}
\end{equation}
with
\begin{equation}
{\cal X}(t_{*}) = \frac{B^2_{0}(1 - \mu_{||})}{4 \mu_{||}}
\exp{\left[- \frac{(3\mu_{||}-1)B_{0} + 2 \mu_{||} M(t_0)
}{(\mu_{||}-1) B_{0}} \right]} < 0 \,, \label{12tstar2}
\end{equation}
and then increases and tends to zero asymptotically. For the
diamagnetic medium we have to change the signs of inequality in
(\ref{12Millustra}) and in (\ref{12tstar2}), i.e., the function
${\cal X}(t)$ reaches the maximum.

\subsection{Third example of cosmological evolution: hidden induction}

Consider the special case when ${\cal X}(t)=0$ despite the
magnetic field is non-vanishing (see, also subsubsection 4.1.4).
Assuming that $P_{||}(t)= 0$,  we guarantee that the first
equation in (\ref{1Plong}) is identically satisfied. Likewise,
assume that $P_{({\rm tr})} = (\gamma -1)W$. Then, Einstein's
field equations (\ref{Ein11})-(\ref{Ein33}) effectively reduce to
the pair of equations
\begin{equation}
\frac{\dot{c}}{c} = H_0 + \frac{\kappa}{2H_0} W \,, \quad
\dot{W} + W H_0 (2\gamma+1) + \frac{\kappa}{2H_0} W^2 = 0 \,.
\label{1e2}
\end{equation}
The solution to the second equation is
\begin{equation}
W(t) = W(t_0) e^{- (2\gamma+1)H_0(t-t_0)}
\left\{1  + \frac{\kappa W(t_0)}{2H^2_0(2\gamma+1)}
\left[1 - e^{- (2\gamma+1)H_0(t-t_0)} \right] \right\}^{-1}
\,,
\label{12Wt}
\end{equation}
thus,
\begin{equation}
c(t) = c(t_0) e^{H_0(t-t_0)}
\left\{1 + \frac{\kappa W(t_0)}{2H^2_0(2\gamma+1)} \left[1 - e^{- (2\gamma+1)H_0(t-t_0)} \right]
\right\}
\,.
\label{12ct1}
\end{equation}
When $t \to \infty$ one has the asymptotic relationship
\begin{equation}
\frac{d}{dt}\left[\log \frac{c(t)}{a(t)}\right] =
\left(\frac{\dot{c}}{c} - \frac{\dot{a}}{a}\right) = \frac{\kappa
W(t_0)}{2H_0} \left[e^{ (2\gamma+1)H_0(t-t_0)} - 1 \right]^{-1}
\rightarrow 0 \,. \label{isotropa}
\end{equation}
This means that the Universe isotropizes.

\subsection{Fourth example of cosmological evolution}

Consider now the special case when ${\cal P}_{||} \equiv
P_{||}(t)- {\cal X}(t) = 0$, $a(t) = a(t_0) \exp\{ H_0(t-t_0)\}$,
$\Lambda = 3 H^2_0$, and $P_{({\rm tr})} = \omega P_{||}$. For
such a model the third Einstein equation (\ref{Ein33}) converts
into identity, the second Einstein equation (\ref{Ein22})
transforms into the equation for $c(t)$
\begin{equation}
\ddot{c} + H_0 \dot{c} + c \left[ \kappa (\omega +1)X(t) - 2 H^2_0 \right] = 0 \,,
\label{13ceq}
\end{equation}
and the first  one, (\ref{Ein11}), gives $W(t)$ if $c(t)$ is
known. Note, that when $\omega=0$ we have ``transversal material
dust", when $\omega=1$, the pressure of matter is isotropic. When
$\omega=-1$, one obtains that ${\cal P}_{({\rm tr})} \equiv
P_{({\rm tr})} {+} {\cal X} {=} {\cal P}_{||} {=} 0$.

\subsubsection{First special case $\omega = -1$}

This case is the simplest, the solution of (\ref{13ceq}) is
\begin{equation}
c(t) = \frac{1}{3} \left[ 2 c(t_0) + \frac{\dot{c(t_0)}}{H_0} \right] e^{H_0(t-t_0)} +
\frac{1}{3} \left[c(t_0) - \frac{\dot{c(t_0)}}{H_0} \right] e^{-2H_0(t-t_0)}
\,.
\label{121c}
\end{equation}
The energy density can be found from the formula
\begin{equation}
\kappa (W + {\cal X}) = - 6 H^2_0 \left\{1  + \left[ \frac{2
c(t_0) + \frac{\dot{c}(t_0)}{H_0}}{c(t_0) -
\frac{\dot{c}(t_0)}{H_0}}\right] e^{3H_0(t-t_0)} \right\}^{-1} \,,
\label{121W}
\end{equation}
where ${\cal X}$ is given by (\ref{12X}). Initial value
$\dot{c}(t_0)$ is connected with $c(t_0)$, $W_0$ and ${\cal X}_0$
by the relation $\frac{\dot{c}(t_0)}{c(t_0)} = H_0 + \frac{\kappa
(W_0+{\cal X}_0)}{2H_0}$, which is a direct consequence of
(\ref{Ein11}). The Universe asymptotically isotropizes, i.e., $a
\propto \exp(H_0 t)$, $c \propto \exp(H_0 t)$, and the total
energy density $ {\cal W} \equiv W+{\cal X}$ decreases as
$\exp(-3H_0 t)$.

\subsubsection{Second special case ($\omega \neq -1$)}

For the special choice of the initial parameter $M(t_0)$, which
yields
\begin{equation}
M(t_0) = \frac{B_{0} (\mu_{||}-1)}{\mu_{||} (2H_0 \tau_0 -1)} \,,
\quad  {\cal X}(t) = \frac{B^2_{0} (2H_0
\tau_0\mu_{||}-1)}{2\mu_{||} (2H_0 \tau_0 -1)} e^{- 4H_0(t-t_0)}
\,, \label{13M0}
\end{equation}
the substitution
\begin{equation}
z= A e^{- 2H_0(t-t_0)} \,, \quad c(t) = z^{\frac{1}{4}} Z(z) \,,
\label{13substi}
\end{equation}
with
\begin{equation}
A \equiv \left\{  \frac{\kappa (\omega+1)B^2_{0} (2H_0 \tau_0\mu_{||}-1)}
{8 H^2_0 \mu_{||} (2H_0 \tau_0 -1)}  \right\}^{\frac{1}{2}}
\,,
\label{13A1}
\end{equation}
reduces the equation (\ref{13ceq}) to
\begin{equation}
z^2 \frac{d^2}{dz^2} Z + z \frac{d}{dz} Z + \left(z^2 - \frac{9}{16}\right)Z = 0
\,.
\label{bessel}
\end{equation}
It is the Bessel equation (see, \cite{AbSt}, Eq. (9.1.1)). The
solution can be expressed in terms of Bessel functions of the real
argument when
\begin{equation}
\tau_0 > \frac{1}{2H_0}  \ \ {\rm or} \ \ \tau_0 < \frac{1}{2H_0 \mu_{||}}
\,,
\label{besselreal}
\end{equation}
for the paramagnetic medium, and when
\begin{equation}
\tau_0 < \frac{1}{2H_0}  \ \ {\rm or} \ \ \tau_0 > \frac{1}{2H_0 \mu_{||}}
\,,
\label{besselreal2}
\end{equation}
for the diamagnetic medium. This equation can also be reduced to
the generalized Bessel equation for the imaginary argument $iz$
(see, \cite{AbSt}, Eq. (9.6.1)), when
\begin{equation}
\frac{1}{2H_0 \mu_{||}}  < \tau_0 <  \frac{1}{2H_0} \,,
\label{besselimag}
\end{equation}
for $\mu_{||} > 1$, and when
\begin{equation}
\frac{1}{2H_0 \mu_{||}}  > \tau_0 >  \frac{1}{2H_0} \,,
\label{besselimag2}
\end{equation}
for $\mu_{||} < 1$. In the last case $c(t)$ can be expressed in
terms of Bessel functions $I_{\nu}(z) \equiv i^{-\nu} J_{\nu}
(iz)$. For simplicity we assume that $\omega +1 >0$ and $\tau_0$
is positive. Then the solution of (\ref{bessel}) is
\begin{equation}
c(t) =  e^{- \frac{1}{2} H_0 (t-t_0)}
\left[ C_1 J_{\frac{3}{4}}\left(A e^{-2H_0(t-t_0)}\right) +
C_2 J_{-\frac{3}{4}}\left(A e^{-2H_0(t-t_0)}\right)
\right]
\,,
\label{13solforc}
\end{equation}
where $J_{\frac{3}{4}}(x)$ and $J_{-\frac{3}{4}}(x)$ are the Bessel functions of
the indices $\nu = \frac{3}{4}$ and $\nu = - \frac{3}{4}$, respectively.
The constants $C_1$ and $C_2$ can be expressed in terms of initial data $c(t_0)$ and
$\dot{c}(t_0)$:
\begin{equation}
C_1 = - \frac{\pi}{2\sqrt{2}} \left[ 4 c(t_0) A J^{ \ '}_{-\frac{3}{4}}(A) +
\left(c(t_0) + \frac{2}{H_0} \dot{c}(t_0) \right) J_{-\frac{3}{4}}(A) \right] \,,
\label{C1}
\end{equation}
\begin{equation}
C_2 = \frac{\pi}{2\sqrt{2}} \left[ 4 c(t_0) A J^{ \ '}_{\frac{3}{4}}(A) +
\left(c(t_0) + \frac{2}{H_0} \dot{c}(t_0) \right) J_{\frac{3}{4}}(A) \right] \,.
\label{C2}
\end{equation}
When $t \to \infty$ the argument of the Bessel functions in (\ref{13solforc}) tends to zero,
and we have the following asymptotic expression
\begin{equation}
c(t \to \infty) = C_2 \frac{2^{\frac{3}{4}}}{\Gamma \left(\frac{1}{4}\right) A^{\frac{3}{4}}}
e^{H_0 (t-t_0)}  \,,
\label{13ctinfty}
\end{equation}
where $\Gamma \left(\frac{1}{4}\right)$ is the Gamma-function.
Thus, the de Sitter regime appears at $t \to \infty$. It is interesting that at $t \to - \infty$,
when the argument of the Bessel functions tends to infinity,
the corresponding formula
\begin{equation}
c(t) \to \frac{1}{A^{\frac{1}{2}}} \sqrt{\frac{2}{\pi}} e^{H_0
(t{-}t_0)} \left[ C_1 \cos{\left(A e^{{-}2H_0 (t{-}t_0)} {-}
\frac{5\pi}{8} \right)} {+} C_2 \sin{\left(A e^{{-}2H_0 (t{-}t_0)}
{+} \frac{\pi}{8} \right)} \right] \label{13ctminusinfty}
\end{equation}
demonstrates the fast quasi-harmonic oscillations of $c(t)$ with
the standard exponential damping. Note that $M(t)$ vanishes when
$M(t_0)= 0$ and $\mu_{||}=1$. This case can also be described by
the formulas (\ref{13solforc}), (\ref{C1}) and (\ref{C2}) with
\begin{equation}
A = \left\{ \frac{\kappa (\omega+1)B^2_{0}}
{8 H^2_0} \right\}^{\frac{1}{2}}
\label{AzeroM}
\end{equation}
and $\tau_0 \neq \frac{1}{2H_0}$. The energy density $W$ as a
solution of (\ref{Ein11}) inherits the dependence on time via the
Bessel functions, we do not reproduce this expression here.

\subsubsection{Third special case}

\noindent When $2H_0 \mu_{||}\tau_0 = 1$, the term ${\cal X}(t)$
reads
\begin{equation}
X(t) = \frac{B^2_{0}}{2} \left[ 1 + \frac{M_0}{B_{0}} \right]
e^{- 2H_0 (1+\mu_{||})(t-t_0)} \,.
\label{13Xspecial2}
\end{equation}
In this case the solution of the equation (\ref{13ceq}) can also be represented in terms of
Bessel functions
\begin{equation}
c(t) = e^{- \frac{1}{2} H_0 (t-t_0)}
\left[ C_1 J_{\frac{3}{4}}\left(\tilde{A} e^{-H_0 (1+\mu_{||})(t-t_0)}\right) +
C_2 J_{-\frac{3}{4}}\left(\tilde{A} e^{-H_0 (1+\mu_{||})(t-t_0)}\right)
\right]
\,,
\label{13cbessel2}
\end{equation}
where
\begin{equation}
\tilde{A} = \left\{  \frac{\kappa (\omega+1)B^2_{0}}
{2 H^2_0 (1+ \mu_{||})^2} \left[ 1 + \frac{M(t_0)}{B_{0}} \right]
\right\}^{\frac{1}{2}}
\,.
\label{13Aspecial2}
\end{equation}
Note that in the ferromagnetic phase $\mu_{||} >> 1$ the argument of Bessel function in
(\ref{13cbessel2}) tends to zero much faster than in case (\ref{13solforc}), i.e., the
isotropization in the ferromagnetic phase takes place faster.

\subsection{Fifth example of cosmological dynamics: \\ non-homogeneous and non-linear
equation of state}

Consider now the special type  of equation of state
\begin{equation}
{\cal P}_{({\rm tr})} = (\gamma -1) {\cal W} + \lambda {\cal X}  -
\frac{\kappa}{4H^2_0} {\cal W}^2 \,. \label{14state}
\end{equation}
Numerous non-homogeneous and non-linear equations of state of such
kind are under discussion (see, e.g., \cite{Odintsov3,Odintsov4}).
As before, we assume that $\Lambda = 3 H^2_0$ and $a(t)= a(t_0)
e^{H_0(t-t_0)}$. For the equation of state (\ref{14state}) the
equation (\ref{conserva}) yields
\begin{equation}
\frac{d}{dt} (W+{\cal X}) + H_0(2\gamma+1)(W+{\cal X}) = - 2H_0
\lambda {\cal X} \,, \label{14conserva}
\end{equation}
whose solution reads
\begin{equation}
W(t) = - {\cal X}(t) + \tilde{L}_1 \ e^{- H_0(2\gamma+1)(t-t_0)} +
\tilde{L}_2 \ e^{- 4H_0(t-t_0)} + \tilde{L}_3 \ e^{-
(2H_0+\frac{1}{\tau_0})(t-t_0)}\,. \label{14W}
\end{equation}
Here the constants $\tilde{L}_1$, $\tilde{L}_2$ and $\tilde{L}_3$ are given by
$$
\tilde{L}_1 \equiv
W(t_0) + \frac{B^2_{0}}{2} + \frac{B_{0} M(t_0)}{2} \
\frac{\left[H_0 \tau_0 (2 \gamma + 2 \lambda -1) -1 \right]}{\left[H_0 \tau_0
(2 \gamma -1) -1 \right]}
$$
\begin{equation}
+
\frac{\lambda B^2_{0}}{\mu_{||} (2\gamma -3)}
\ \frac{\left[H_0 \tau_0 \mu_{||}(2 \gamma -1) -1 \right]}{\left[H_0 \tau_0
(2 \gamma -1) -1 \right]}
\,,
\label{14L1}
\end{equation}
\begin{equation}
\tilde{L}_2 \equiv -
\frac{\lambda B^2_{0}}{\mu_{||}(2\gamma -3)}
\ \frac{\left[2H_0 \tau_0 \mu_{||} -1 \right]}{(2H_0 \tau_0 -1 )}
\,,
\label{14L2}
\end{equation}
\begin{equation}
\tilde{L}_3 \equiv
- B_{0} M(t_0) \
\frac{H_0 \lambda \tau_0}{\left[H_0 \tau_0 (2 \gamma -1) -1 \right]} +
\frac{\lambda B^2_{0}}{\mu_{||}(2H_0 \tau_0 -1)}
\ \frac{\left[H_0 \tau_0 (\mu_{||} -1) \right]}{\left[H_0 \tau_0 (2 \gamma -1) -1 \right]}
\,.
\label{14L3}
\end{equation}
Likewise, for $c(t)$ we obtain
$$
\log{\left( \frac{c(t)}{c(t_0)}\right)} = H_0 (t-t_0)
+ \frac{\kappa \tilde{L}_1}{2H^2_0(2\gamma+1)} \left[ 1 - e^{- H_0(2\gamma+1)(t-t_0)} \right]
$$
\begin{equation}
+ \frac{\kappa \tilde{L}_2}{8H^2_0} \left[ 1 - e^{- 4 H_0 (t-t_0)} \right]
+ \frac{\kappa \tilde{L}_3}{2H_0(2H_0+\frac{1}{\tau_0})}
\left[ 1 - e^{- (2H_0+\frac{1}{\tau_0})(t-t_0)} \right]
\,.
\label{14c}
\end{equation}
In the asymptotic regime at $t \to \infty$ one obtains $c(t) \to c(\infty) e^{H_0 (t-t_0)}$, where
\begin{equation}
c(\infty) \equiv c(t_0) \
\exp\left\{\frac{\kappa}{2H_0}\left[\frac{\tilde{L}_1}{H_0(2\gamma {+}1)} {+}
\frac{\tilde{L}_2}{4H_0}
+ \frac{\tilde{L}_3 \tau}{(2H_0 \tau_0 {+} 1)} \right]\right\}\,.
\label{14casympta}
\end{equation}
Thus, the isotropization takes place, as it should.
For $t \to - \infty$,  $c(t)$ decreases superexponentially.

\section{Discussion}

We have considered the simplest model of the one-dimensional
relaxation of matter magnetization in a strong magnetic field in
the framework of the extended Einstein-Maxwell theory applied to
Bianchi-I cosmological model. We have shown that this model admits
a set of exact analytical solutions, depending on the set of
guiding parameters. Let us emphasize the main aspects of the
obtained results.

\vspace{2mm}

\noindent{\it 1. Analogy with extended (causal) thermodynamics.}

\noindent The key element of the extended Einstein - Maxwell
theory in the context of anisotropic Bianchi-I model is the
one-dimensional relaxation equation (\ref{CEm}) for the
magnetization $M(t)$. The key element of the extended irreversible
thermodynamics in the context of isotropic Friedmann model is a
relaxation equation for the bulk viscous pressure $\sigma(t)$
(see, e.g., \cite{Pavon91}). These two equations
\begin{equation}
\tau \dot{M} + M = \left(\frac{1}{\mu_{||}} - 1 \right) B_0 \left(\frac{a(t_0)}{a(t)}\right)^2
 \ \ \ {\rm and} \ \
\ \tau \dot{\sigma} + \sigma = - 3 \zeta \frac{\dot{a}}{a}
\label{analog}
\end{equation}
look similar. ($\zeta$ is a bulk viscosity coefficient). In both
cases the rate of evolution of the scale factor $a(t)$
predetermines the relaxation properties of $M(t)$ or $\sigma(t)$.
In both cases the function $M(t)$ or $\sigma(t)$, appears in the
right-hand-side of the Einstein equations as an element of the
source term. In both models new degrees of freedom, activated in
matter by the cosmological evolution, change the rate of
expansion. In order to check this claim one can simply compare the
results obtained for $M(t)=0$ and $M(t) \neq 0$. As an example,
let us compare the expressions (\ref{WW}), (\ref{HHH1}),
(\ref{HHH2}) and (\ref{caH}) with those at $M_0 = 0$,
$\mu_{||}=1$, $\xi =0$ (i.e., at $K_2=0$). The difference is that
the magnetization adds a principally new second term in
(\ref{WW}), describing the evolution of the energy density scalar,
as well as the new last term in (\ref{HHH1}), describing the rate
of expansion in the cross-section $x^1Ox^2$. Taking into account
the formula (\ref{caH}), one can see that the modifications in
$H_{(a)}(t)$ lead to the changes in the rate of evolution in the
direction $x^3$. Moreover, the presence of the new terms $x^{-2-
\frac{1}{\xi}}$, allows us to choose the phenomenological
parameter $\xi$ so that this term becomes of the leading order at
$x \to \infty$ in comparison with the terms $x^{-3}$, $x^{-4}$ and
$x^{-2\gamma}$ in the formula (\ref{HHH1}). In such a case just
the magnetization predetermines the rate of cosmological evolution
at $t \to \infty$, and the relaxation time $\tau(t) = \xi
H_{a}(t)$ introduces a new expansion time scale.

In both theories  the relaxation time $\tau$ is considered to be a function of cosmological
time $t$ and is a subject of modeling. In the extended irreversible thermodynamics the relaxation
time is considered as $\tau = \frac{\zeta}{W}$, where $\zeta = \alpha W^q$ (in our definition of the
energy density scalar) (see, e.g., \cite{Pavon91} - \cite{Maartens95}). When the function $W(t)$ is
obtained from the cosmological dynamics, the function $\tau(W(t))$ becomes an alternative
representation of the function $\tau(H(t))$.

The main difference of the results is connected with the fact that
the Bianchi-I model is anisotropic, and $M(t)$ is in fact a
projection of the magnetization on the direction pointed by the
magnetic field. In the isotropic Friedmann model $\sigma$ is a
scalar describing the isotropic bulk viscous pressure. As a
consequence, $H(t)$ in the Friedmann model is positive and the
right-hand-side of the relaxation equation for $\sigma$ is always
negative. The sign of the right-hand-side of the relaxation
equation for $M(t)$ depends on the sign of $B_0$, as well as on
the sign of the difference $(\mu_{||} -1)$. Respectively, the
obtained magnetization may be positive or negative depending on
the (random) initial value $M_0$, relaxation time and magnetic
permeability. This option allows to consider a principally new
situation, when magnetic field and magnetization are
non-vanishing, nevertheless, the total magnetic source term ${\cal
X} = \frac{1}{2} B(B+M)$ is equal to zero and disappears from the
Einstein equations. Such solutions are discussed in the
subsections 4.1.4.  and 5.1.2.

\vspace{2mm}

\noindent{\it 2. Monotonic and non-monotonic expansion}

\noindent Classical models with pure magnetic field are
characterized by the non-negative source term ${\cal X} =
\frac{1}{2} B^2 \geq 0$, which decreases monotonically as
$a^{-4}$. The magnetization changes the situation: ${\cal X}$ may
be positive, negative or equal to zero. Generally, ${\cal X}(t)$
is not monotonic function any longer, it may possess one, two or
more extremums. As a consequence of this behaviour, the function
$H_{(a)}(t)$ is not monotonic, thus, in the evolution of the
Universe in the cross-section $x^1Ox^2$ there are periods of
(transversal) acceleration and deceleration. The simplest
behaviour $H_{(a)}(t)$ is characterized by  the presence  of one
minimum or maximum, and by the asymptotic de Sitter regime $H_a(t)
\to \sqrt{\frac{\Lambda}{3}}$ (see, e.g., subsection 5.2.3). The
behaviour of $c(t)$ is also non-monotonic in this case. More
complicated situation is characterizes by the solution for $c(t)$,
presented in terms of Bessel functions (see, subsubsection 6.3.2).
The function $c(t)$ behaves quasi-periodically, and one can expect
that the number of periods of the longitudinal acceleration and
deceleration is infinite.

\vspace{2mm}

\noindent{\it 3. Guiding and resonance parameters}

\noindent The considered extended Einstein-Maxwell model is
characterized by eight guiding parameters: $\xi$ or $\tau_0$,
$\mu_{||} - 1$, $B_0$, $M_0$, $W_0$, $\gamma$, $H_{(a)}(t_0)$ and
$\Lambda$. There are several underlined values of the parameter
$\xi$ (in the model of variable relaxation time) and of the
parameter $\tau_0$ (in the model of constant relaxation time). The
values $\xi = \frac{1}{2}$ and $\frac{1}{\tau_0}= 2 H_0$,
respectively, are in fact resonance parameters, which appear in
the integration of the differential equation for $M(t)$. In such a
resonance case the relaxation time $\tau = \frac{1}{2}
H^{-1}_{(a)}(t)$ or $\tau_0 = \frac{1}{2}H^{-1}_0$ coincides with
the characteristic time of the evolution of the function $B(t) =
F_{12} \ a^{-2}(t)$, which provides the dynamics of magnetization.
In case of resonance the function $x^{-\frac{1}{\xi}}$, as a part
of $M(t)$ (\ref{1M1}), has to be replaced by $x^{-2} \log x$ (see,
(\ref{1Mlog})). It is very interesting to emphasize that in
\cite{Maartens95} the special value of the parameter $q$, $q=
\frac{1}{2}$, leads to the law $\tau \sim H^{-1}$ for the
relaxation of the bulk viscosity pressure. The value $\xi = 1$ is
evidently the resonance value of the parameter $\xi$, but it has
another origin. When $\xi = 1$ the rate of change of the function
$H_{(a)}(t)$ coincides with that of ${\cal X}(t)$. As a
consequence, the function $x^{-\left( 2 + \frac{1}{\xi} \right)}$
in (\ref{111H}) has to be replaced by $x^{-3} \log x$ (see,
(\ref{11H1})). The special values $\frac{1}{\xi} = 2 (\gamma -1)$,
$\frac{1}{\tau_0} = H_0 (2 \gamma -1)$ (see,
(\ref{WW}))-(\ref{HHH2})) and (\ref{14L1}))-(\ref{14L3})) can also
be considered as some analogs of the relation $\tau^{-1} = 3H_0
\frac{(2 - \epsilon \gamma)}{2\gamma}$ appeared in
\cite{Maartens95} in the context of evolution of the bulk viscous
pressure. The special value $\gamma = \frac{3}{2}$, appearing in
(\ref{HHH2}), relates to the vanishing trace of the matter
pressure tensor $P_{1}+P_{2}+P_{3}=2P_{({\rm tr})} + P_{||} = 0$.
The special value $\frac{1}{\tau_0} = 2 H_0 \mu_{||}$ appears in
the context of the vanishing ${\cal X}(t)$ (see, (\ref{12zeroX})).
Finally, the model characterized by the constant values of
$P_{||}$, $P_{({\rm tr})}$, $W$ and ${\cal X}$, discussed in the
subsubsection 5.2.1 in context of the special condition
$H_{(a)}(t_0) =0$, also has an appropriate analog in
\cite{Maartens95}.

\vspace{2mm}

\noindent{\it 4. Isotropization}

\noindent The models, in which the cosmological constant $\Lambda$
is non-vanishing, isotropize at $t \to  \infty$. The exceptional
case (see, (\ref{sigma112})) corresponds to $\Lambda = 0$. The
first novelty of the obtained results is that at $\xi = -
\frac{1}{2}$ the magnetized matter can effectively redefine the
cosmological constant (see, (\ref{cc})). This case is exotic,
since negative $\xi$ corresponds to magnetic instability and the
magnetization increases with time. Nevertheless, when $\Lambda
=0$, this effect can in principle produce a non-vanishing
effective cosmological constant. The second novelty is connected
with the non-monotonic character of the isotropization.

\vspace{3mm}
\noindent
{\bf Acknowledgments}

\noindent
\noindent
This work was supported by the Catalonian Government (grant 2004 PIV1 35) and was done
in the Department of Physics of the Universidad Aut\'onoma de Barcelona.
The author is grateful to the colleagues of Group of Statistical Physics for hospitality.
The author is especially thankful to Prof. D. Pav\'on for reading the manuscript, helpful
discussions and advices.

\section{Appendix: Variation of the tetrad vectors}

Let $X^i_{(a)}$ be the set of four tetrad four-vectors, whose
index $(a)$ runs over four values: $(0),(1),(2),(3)$. Let
$X^i_{(0)}$ coincide with $U^i$, the four-vector of velocity of
the medium as a whole, and let $X^i_{(3)}$ coincide with $X^i$,
the director four-vector of the medium with uni-axial symmetry.
The tetrad four-vectors are assumed to satisfy the orthogonality -
normalization rules
\begin{equation}
g_{ik} X^i_{(a)}X^k_{(b)} = \eta_{(a)(b)} \,, \quad  \eta^{(a)(b)}
X^p_{(a)}X^q_{(b)} = g^{pq} \,, \label{tetra111}
\end{equation}
where $\eta_{(a)(b)}$ denotes the Minkowski matrix, diagonal
$(1,{-}1,{-}1,{-}1)$. Since the tetrad four-vectors are linked by
the relation containing the metric,  we have to define the formula
for the variation $\frac{\delta X^i_{(a)}}{\delta g^{pq}}$.
Varying the first and second relations (\ref{tetra111}) with
respect to the metric, we obtain, respectively,
\begin{equation}
X_{k(b)} \delta X^k_{(a)} + X_{k(a)} \delta X^k_{(b)}  = -
X^i_{(a)}X^k_{(b)} \delta g_{ik} \,, \label{tetra2}
\end{equation}
\begin{equation}
\delta g^{pq} = \eta^{(a)(b)} \left[ X^q_{(b)} \delta X^p_{(a)} +
X^p_{(a)} \delta X^q_{(b)} \right]  \,. \label{tetra3}
\end{equation}
The variation of arbitrary origin $\delta X^i_{(a)}$ (not
necessarily caused by the metric variation) can be represented as
a linear combination  of the tetrad  four-vectors:
\begin{equation}
\delta X^i_{(a)} = X^i_{(c)} Y^{ \ \ (c)}_{(a)} \,. \label{tetra5}
\end{equation}
The tetrad tensor $Y^{ \ \ (c)}_{(a)}$ is not generally symmetric.
Using the convolution of (\ref{tetra3}) with tetrad vectors, we
obtain
\begin{equation}
Y^{(a)(b)} + Y^{(b)(a)} = \delta g^{pq} X_{p}^{(a)}X_{q}^{(b)} \,,
\label{tetra6}
\end{equation}
where we use standard rules for the indices, e.g., $X_{q}^{(f)} =
\eta^{(f)(b)} g_{qm} X^m_{(b)}$. Consequently, the symmetric part
of the quantity $Y^{(a)(b)}$, indicated as $Z^{(a)(b)}$, can be
readily found:
\begin{equation}
Z^{(a)(b)} = \frac{1}{2} \delta g^{pq} X_{p}^{(a)}X_{q}^{(b)} \,,
\label{tetra7}
\end{equation}
and the law (\ref{tetra5}) reads now
\begin{equation}
\delta X^i_{(a)} = \frac{1}{4} \delta g^{pq} \left[X_{p (a)}
\delta^i_q + X_{q (a)} \delta^i_p \right] + X^i_{(c)} {\cal
Z}_{(a)}^{\ (c)} \,. \label{tetra8}
\end{equation}
Here ${\cal Z}_{(a)}^{\ (c)}$ is a skew-symmetric part of
$Y_{(a)}^{(c)}$, i.e., $2 {\cal Z}_{(a)(c)} \equiv Y_{(a)(c)} -
Y_{(c)(a)}$.  Therefore, the variation of the metric produces the
variation of the tetrad, described by (\ref{tetra8}) with
vanishing skew-symmetric part ${\cal Z}_{(a)(c)}$. Thus, one
finally has
\begin{equation}
\frac{\delta X^i_{(a)}}{\delta g^{pq}} = \frac{1}{4}  \left[X_{p
(a)} \delta^i_q + X_{q (a)} \delta^i_p \right] \,, \label{tetra9}
\end{equation}
and we can use this formula for the variation of the four-velocity
vector $U^i \equiv X^i_{(0)}$ and for the variation of the
space-like vector $X^i\equiv X^i_{(3)}$.

\end{document}